\documentclass[aps,prd,twocolumn,nofootinbib,showpacs,groupedaddress]{revtex4-1}

\usepackage{amsfonts}
\usepackage{amsmath}
\usepackage{amssymb}
\usepackage{bm}
\usepackage{dcolumn}
\usepackage{graphicx}
\usepackage{graphics}
\usepackage[latin1]{inputenc}
\usepackage{latexsym}
\usepackage{rotating}
\usepackage{hyperref}
\usepackage[all]{hypcap}
\usepackage{xspace} % Sensible space treatment at end of simple macros
\usepackage[usenames]{color}
\usepackage{mathrsfs}

\usepackage{ulem}
\definecolor {darkgreen}{rgb}{0.2,0.7,0.2}

\newcommand{\be}{\begin{equation}}
\newcommand\ee{\end{equation}}

\newcommand\bw{\begin{widetext}}
\newcommand\ew{\end{widetext}}

\newcommand{\nn}{\nonumber}

\newcommand{\SN}{{\mbox{\tiny sn}}}
\newcommand{\SE}{{\mbox{\tiny se}}}
\newcommand{\LL}{{\mbox{\tiny l}}}
\newcommand{\B}{{\mbox{\tiny b}}}
\newcommand{\GR}{{\mbox{\tiny GR}}}
\newcommand{\EA}{\mbox{\tiny EA}}
\newcommand{\sr}{{\mbox{\tiny sr}}}
\newcommand{\fr}{\mbox{\tiny fr}}
\newcommand{\ppE}{\mbox{\tiny ppE}}
\newcommand{\ppN}{\mbox{\tiny ppN}}
\newcommand{\PN}{\mbox{\tiny PN}}
\newcommand{\Newt}{\mbox{\tiny Newt}}
\newcommand{\KG}{\mbox{\tiny KG}}
\newcommand{\MG}{\mbox{\tiny MG}}

\newcommand{\beaz}{\tilde{\beta}_{0\PN, \EA}}
\newcommand{\bkgo}{\tilde{\beta}_{-1\PN, \KG}}
\newcommand{\bkgz}{\tilde{\beta}_{0\PN, \KG}}

\begin{document}
\title{Projected Constraints on Lorentz-Violating Gravity with Gravitational Waves}

\author{Devin Hansen}
\author{Nicol\'as Yunes}
\author{Kent Yagi}
\affiliation{Department of Physics, Montana State University, Bozeman, MT 59717, USA.}

\date{\today}

%%%%%%%%%%%%%%%%%%%%%%%%%%%%%%%%%%%%%%%%%%%%%%%%%
\begin{abstract} 

Gravitational waves are excellent tools to probe the foundations of General Relativity in the strongly dynamical and non-linear regime. 
One such foundation is Lorentz symmetry, which can be broken in the gravitational sector by the existence of a preferred time direction, and thus, a preferred frame at each spacetime point.
This leads to a modification in the orbital decay rate of binary systems, and also in the generation and chirping of their associated gravitational waves.   
We here study whether waves emitted in the late, quasi-circular inspiral of non-spinning, neutron star binaries can place competitive constraints on two proxies of gravitational Lorentz-violation: Einstein-\AE{}ther theory and khronometric gravity.  
We model the waves in the small-coupling (or decoupling) limit and in the post-Newtonian approximation, by perturbatively solving the field equations in small deformations from General Relativity and in the small-velocity/weak-gravity approximation. 
We assume a gravitational wave consistent with General Relativity has been detected with second- and third-generation, ground-based detectors, and with the proposed space-based mission, DECIGO, with and without coincident electromagnetic counterparts.
Without a counterpart, a detection consistent with General Relativity of neutron star binaries can only place competitive constraints on gravitational Lorentz violation when using future, third-generation or space-based instruments. 
On the other hand, a single counterpart is enough to place constraints that are 10 orders of magnitude more stringent than current binary pulsar bounds, even when using second-generation detectors. 
This is because Lorentz violation forces the group velocity of gravitational waves to be different from that of light, and this difference can be very accurately constrained with coincident observations. 

\end{abstract}


\maketitle
\allowdisplaybreaks

%%%%%%%%%%%%%%%%%%%%%%%%%%%%%%%%%%%
\section{Introduction}

\emph{Here be dragons}. This is the warning that medieval cartographers would use to signal a region in their maps that had not yet been explored, to signal the frontiers of human knowledge. One of today's frontiers in physics is the strongly dynamical and non-linear regime of the gravitational interaction. We {\emph{choose}} to believe that General Relativity (GR) is the correct description of Nature in this regime, when black holes collide and neutron stars (NS) merge into each other. This choice, of course, is very sensible, as it is rooted in the fantastic success of GR in describing low energy phenomena, such as physics in the Solar System, and certain strong field physics, such as in binary pulsars~\cite{will-living}. However, assuming that GR is also valid in the dynamical and non-linear regime may be a dangerous extrapolation that must be verified. 

Perhaps one of the most exotic of these ``dragons'', one of the most interesting proposals to modify GR, is the violation of Lorentz symmetry in gravity.  Lorentz symmetry, i.e.~that experimental results are independent of the inertial frame used to carry them out, is a pillar of Special Relativity and many other field theories. Violations of Lorentz symmetry in matter interactions is very well-constrained by observations and experiments~\cite{Mattingly:2005re, Jacobson:2007fh}, but such tests do not strongly constrain Lorentz violations that are dominantly active only in the gravitational sector. Observational constraints on the existence of a preferred gravitational frame are not strong, except perhaps for certain Solar System constraints and the very recent constraints placed with binary pulsars~\cite{Yagi:2013qpa,Yagi:2013ava} and with cosmological observations~\cite{Audren:2014hza}.  The dynamical and non-linear regime has yet to be explored as a possible source for further constraints, and perhaps, experiments that can sample this regime may be the best ``sword'' to slay such dragons.

The Standard Model Extension~\cite{Colladay:1998fq,Kostelecky:1998id,Kostelecky:1999rh,Kostelecky:2003fs,Kostelecky:2008ts,Kostelecky:2010ze} has been proposed as a model-independent way to map various observations to tests of Lorentz invariance. Alternatively, Einstein-\AE{}ther theory~\cite{Jacobson:2000xp, Eling:2004dk} and khronometric gravity~\cite{Blas:2009qj,Blas:2009yd} can be used as representative models of theories that break Lorentz symmetry in gravity~\cite{Yagi:2013qpa,Yagi:2013ava}. Einstein-\AE{}ther theory accomplishes this by coupling the metric to a vector field of everywhere unit magnitude at the level of the action. The latter contains the most generic correction to the Einstein-Hilbert term that has a unit timelike vector and quadratic combinations of its first derivative. Khronometric theory instead contains a globally preferred frame selected by a \textit{khronon} scalar field, which is responsible for globally breaking Lorentz symmetry. Such a theory is realized as the low-energy limit of a UV complete, power-counting renormalizable theory~\cite{Horava:2009uw}. For a more detailed discussion of these theories, see~\cite{Jacobson:2004ts, Foster:2007gr, Mattingly:2005re, Jacobson:2000xp, Eling:2004dk, Blas:2009qj, Yagi:2013ava} 

The strength of Lorentz violations is encoded in the magnitude of the coupling parameters of the theory. Solar System observations have forced Einstein-\AE{}ther theory and khronometric gravity to effectively depend only on two combinations of coupling parameters, $(c_{+},c_{-})$ and $(\beta_{\KG},\lambda_{\KG})$ respectively. Recent constraints with binary pulsar observations~\cite{Yagi:2013qpa,Yagi:2013ava} and cosmological observations~\cite{Audren:2013dwa,Audren:2014hza} have forced these parameters to satisfy $c_{+} \lesssim 0.03$ and $c_{-} \lesssim 0.003$ for Einstein-\AE{}ther theory and $\beta_{\KG} \lesssim 0.005$ and $\lambda_{\KG} \lesssim 0.1$ for khronometric gravity.

Binary pulsar constraints are particularly strong because one of the signatures of gravitational Lorentz violation is the excitation of dipole radiation due to the presence of propagating vector and scalar modes. Dipole radiation modifies the rate of change of the orbital period in binary systems, which radio telescopes have observed to be in agreement with GR to exquisite levels. Thus, binary pulsar observations can strongly constrain the existence of such Lorentz-violating effects, and thus, the magnitude of the coupling parameters of Einstein-\AE{}ther theory and khronometric gravity.

Such orbital decay is an inherently non-linear and dynamical effect, and thus, not just binary pulsars, but rather {\emph{all}} astrophysical observations that can sample this regime may have a shot at constraining gravitational Lorentz violation. One such observation is the detection of gravitational waves with ground-based interferometers, such as Advanced LIGO (aLIGO)~\cite{ligo,Abramovici:1992ah} and Advanced Virgo (aVirgo)\cite{virgo,Caron:1997hu}. These second-generation, gravitational wave detectors will be operating at design sensitivity in the next few years, hopefully detecting from a few to tens of NS and black hole binaries per year with signal-to-noise ratios (SNRs) in the 10s. Third-generation detectors, such as LIGOIII~\cite{Adhikari:2013kya} and the Einstein Telescope (ET)~\cite{et,Punturo:2010zz}, are also being planned for the next decade with the aim to detect hundreds of events per year with SNRs in the 100s. Japan's proposed space-based mission, DECIGO~\cite{decigo,Seto:2001qf}, is expected to detect signals with even larger SNR at lower frequencies.

Gravitational waves and binary pulsar astrophysics are at quite different stages of development. Radio astronomers detected the first signals almost 40 years ago~\cite{Hulse:1974eb} and have continued to monitor these systems ever since. By now, the pulsar community has detected many tens of pulsars in binaries at varying levels of precision. On the other hand, the gravitational wave community has not yet made a detection, with the first ones expected to arrive in the next few years and to be extremely weak. What matters here is that pulsar observations can do data analysis with a combined SNR that is \emph{much} higher than what gravitational wave analysts will have available in the first decade of gravitational wave astrophysics. 

The above suggests that binary pulsars may have a better handle at constraining the existence of Lorentz-violating effects in gravity than gravitational wave observations. Indeed, this has been predicted to be the case for the presence of dipolar radiation, with both a Fisher analysis and Bayesian model-selection tools~\cite{cornishsampson}. Nonetheless, one may wonder (i) whether this generic prediction holds true for other Lorentz-violating effects, such as the modification of the graviton propagation speed and modifications to propagating, quadrupole tensor modes, and (ii) if it does hold true, how far out would gravitational wave constraints be from binary pulsar ones. Once these questions are answered, one could determine whether future, third-generation detectors would be able to place bounds comparable with current binary pulsar ones. 

These questions are the main theme of this paper. We address them by studying gravitational waves in Einstein-\AE{}ther theory and khronometric gravity, emitted by non-spinning NSs in their late, quasi-circular inspiral phase, i.e.~when the gravitational waves emitted are at frequencies higher than $1$--$10$ Hz, or equivalently when their separation is smaller than $10^{3}$ km. We neglect the merger phase, as this occurs at frequencies above $1 {\;} {\rm{kHZ}}$ for binary NSs, where detectors are much less sensitive. We calculate how gravitational Lorentz violating effects propagate into the response function of a detector given an impinging gravitational wave, both in the time and frequency domains. The latter is treated in the stationary phase approximation (SPA), the leading-order term in the method of steepest descent employed to solve generalized Fourier integrals~\cite{bender}. 

Such an analysis is carried out through certain approximations. First, we use a \emph{small-velocity/weak-field approximation} (the so-called post-Newtonian (PN) formalism~\cite{Blanchet:2002av}) to model the orbital dynamics and gravitational wave generation in the inspiral phase. Second, we use a \emph{small-deformation} approximation (the so-called decoupling limit), where Lorentz violating effects are assumed to lead to small corrections to GR predictions. This approximation is justified in Einstein-\AE{}ther theory and khronometric gravity, given the stringent constraints already placed through binary pulsar observations~\cite{Yagi:2013qpa,Yagi:2013ava}. These two approximations make the calculation of the gravitational wave response function analytically tractable.  

% % % % % % % % % % % % % % % % %
\begin{figure*}[ht]
\includegraphics[width=8.75cm,clip=true]{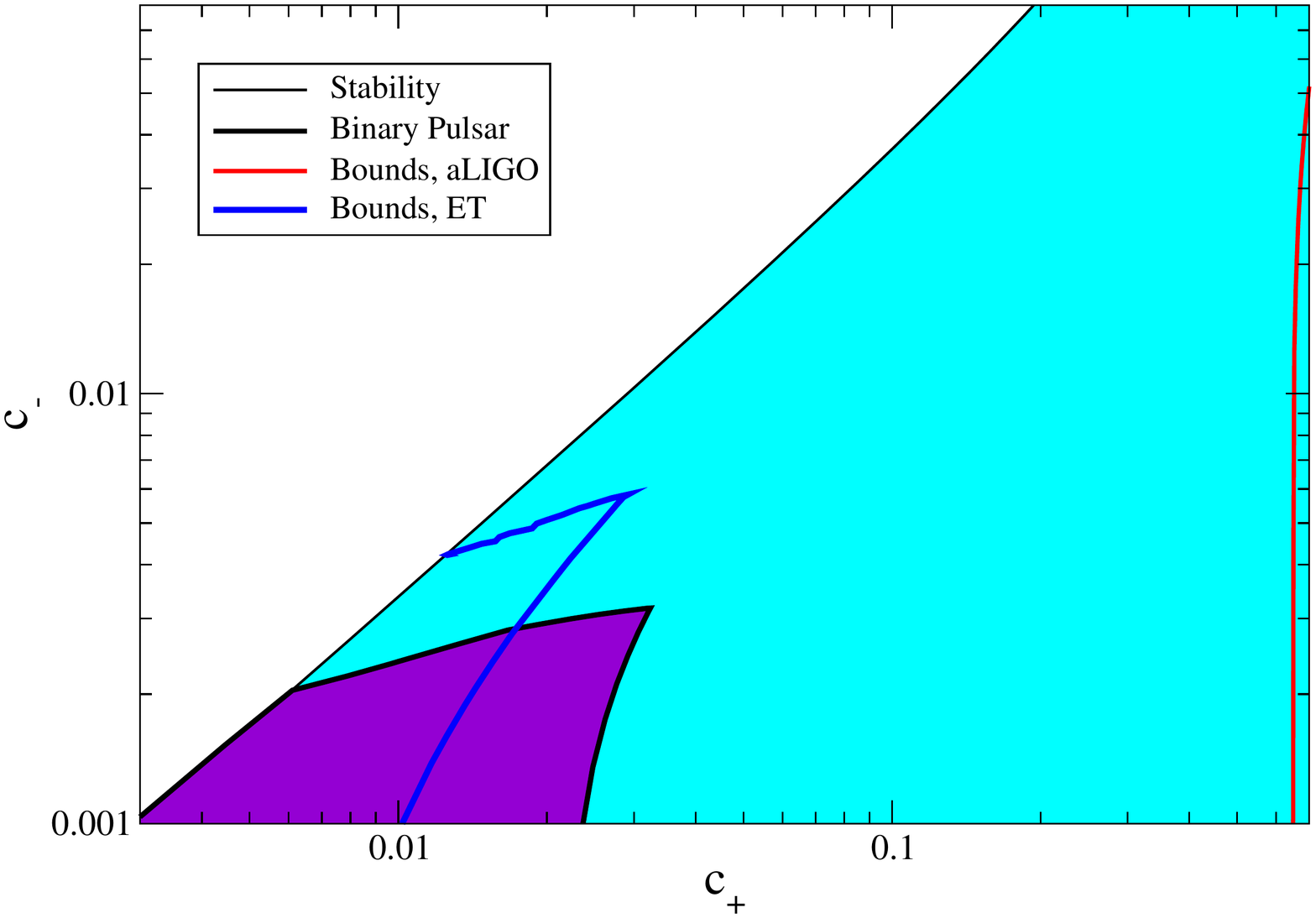} 
\includegraphics[width=8.75cm,clip=true]{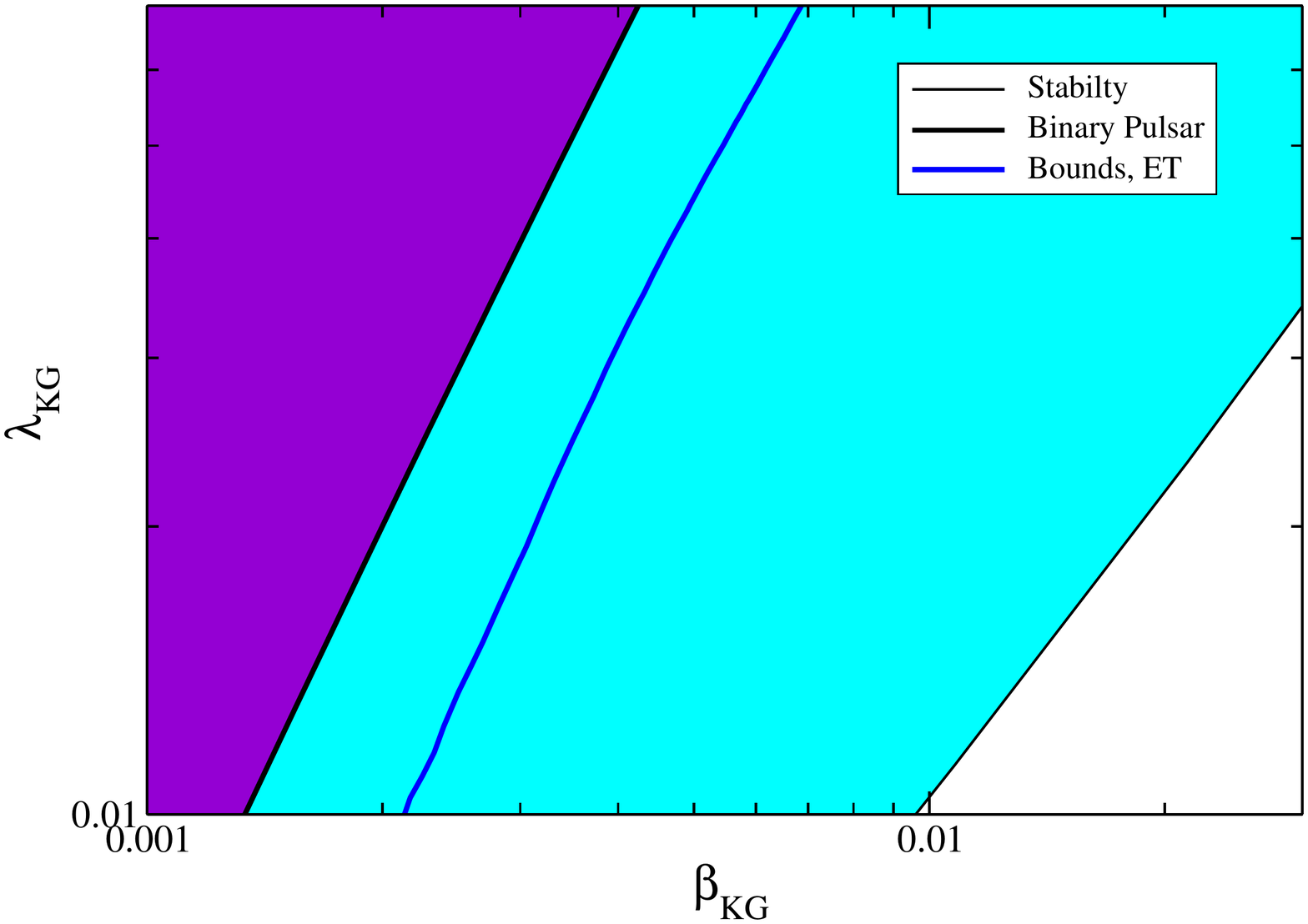}
\caption{\label{fig:Constraints} Projected constraints on $c_+$ and $c_-$ (Left), and $\beta_{\KG}$ and $\lambda_{\KG}$ (Right) assuming a gravitational wave observations from the quasi-circular inspiral of non-spinning NSs consistent with GR, at a luminosity distance of $270 \; {\rm{Mpc}}$. This corresponds to SNRs 10 and 130 for aLIGO and ET respectively, averaging over sky location.  The region below and to the left of the contours are the allowed regions for the respective coupling parameters.  The purple shaded region is the allowed region from binary pulsar and cosmological observations, as calculated in \cite{Yagi:2013qpa,Yagi:2013ava}, and the lighter shaded region is the allowed region from stability.  As a representative case we show here contours produced given a detection with aLIGO and with ET. In the khronometric gravity case, the constraint from aLIGO is outside the scale of the plot.}
\end{figure*}
% % % % % % % % % % % % % % % % % % %

With the response function at hand, we carry out a Fisher analysis~\cite{cutlerflanagan}, assuming a gravitational wave detection consistent with GR with aLIGO (second-generation detector), LIGOIII/ET (third-generation detectors), and the proposed space-based DECIGO mission. 
We find that second-generation detectors alone will not be able to place constraints on gravitational Lorentz violation that are more stringent than current binary pulsar bounds, assuming signals from NS binaries. 
This is because the leading-order modifications to the gravitational wave phase enter at $-1$PN order\footnote{A correction to some leading order expression that is proportional to $v^{2N}$ is said to be of $N$PN order.} due to dipole radiation and at 0PN (or Newtonian order) due to modifications to quadrupole radiation. 
The dipole correction, however, is proportional to the square of the difference of the NS sensitivities (related to the Lorentz-violating field's self-energy), and thus to the square of the NS mass difference, which is small for NS binaries.
The quadrupole correction is partially degenerate with the chirp mass, which also enters at 0PN order in the gravitational wave phase. 
All of this combines to lead to projected constraints with second-generation detectors from NS binaries that are at least 2 orders of magnitude weaker than current binary pulsar bounds. 
Third-generation ground and space-based detectors, however, would be able to place constraints that are comparable, and in some instances, stronger than current bounds by a factor of roughly 2. 
Note that these bounds may not hold for mixed binary systems, however such an analysis would require the calculation of black hole sensitivities, which has yet to be done and is not considered in this paper.

Figure~\ref{fig:Constraints} shows projected and current constraints on the coupling parameters of Einstein-\AE{}ther theory and khronometric gravity. This is an exclusion plot, so the area under (over) the curves are coupling parameter space regions that are still allowed for Einstein-\AE{}ther theory (khronometric gravity). The area bounded by the black thin line (cyan region) and the solid thick line (purple region) are the currently allowed regions after stability constraints and binary pulsar/cosmology constraints respectively~\cite{Yagi:2013qpa,Yagi:2013ava,Audren:2014hza}. The area to the left of the red curve and bounded by blue curves are the new, projected allowed regions after a gravitational wave detection with second-generation and third-generation detectors respectively. Observe that in the Einstein-\AE{}ther case (left panel), ET detections would be required to cut into the current allowed region. In the khronometric gravity case (right panel), the aLIGO constraint is outside the scale of the figure, and we see that not even ET detections would suffice to place constraints comparable to current ones.  

But what if a gravitational wave detection occurs simultaneously with an electromagnetic (e.g.~gamma ray or x-ray) observation? This could, for example, happen if short gamma-ray bursts are produced by the merger of binary NSs, or if a supernovae explosion is detected both in the x-rays and gravitationally. In the former case, one would also need the geometry of the NS remnant to be such that the burst is along Earth's line of sight. In the latter case, one would need the supernovae to occur sufficiently close to Earth for detection. The probability of simultaneously detecting either is not high, but a \emph{single} coincident detection would here suffice to carry out a stringent test of GR. 

Given such a coincident detection, one can then constrain the propagation speed of gravitational radiation relative to the speed of light by measuring the time delay between photon or neutrino arrivals relative to the gravitational wave arrival~\cite{Nishizawa:2014zna}. The mapping between times of arrival and propagation speed requires knowledge of the distance to the source, which would here be provided by the gravitational wave measurement. Even though the latter is not expected to be accurately determined, one can still place stringent constraint on the gravitational wave propagation speed. Since Lorentz-violating theories generally modify this speed, we find that a coincident detection would allow us to constrain $c_{+}$ and $\beta_{\KG}$ 10 orders of magnitude more stringently than current binary pulsar bounds. Although no constraint can be placed on the other coupling parameters of the theory (like $c_{-}$ in Einstein-\AE{}ther theory and $\lambda_{\KG}$ in khronometric gravity), such stringent constraints on $c_{+}$ and $\beta_{\KG}$ greatly restrict the coupling parameter space.

The remainder of the paper deals with the details of this calculation. 
Section~\ref{sec:ABC} gives a brief review of Einstein-\AE{}ther theory and khronometric gravity, including the current bounds placed on the coupling parameters through binary pulsar constraints. 
Section~\ref{sec:Response} constructs the response function for a general modified gravity theory.
Sections~\ref{sec:AE-theory} and~\ref{sec:KG-theory} construct the specific response functions for Einstein-\AE{}ther theory and khronometric gravity, respectively.
Section~\ref{sec:Fisher} carries out a Fisher analysis and predicts the bounds that we could place on these theories, given a gravitational wave detection consistent with GR.
Section~\ref{sec:gwem} calculates the bounds given coincident gravitational wave and electromagnetic signals.
Section~\ref{sec:Conclusions} gives a brief summary and provides some closing remarks.

Henceforth, we follow the conventions of ~\cite{MTW} in which Greek letters signify spacetime indices and Latin letters represent purely spacial indices. We use the ${\rm{diag}}(-1, 1, 1, 1)$ signature and geometric units in which $G_{N}=1=c$. Here, $G_{N}$ is the gravitational constant that enters Newton's third law in the weak-field.

%%%%%%%%%%%%%%%%%%%%%%%%%%%%%%
\section{Review of Lorentz-Violating Theories}
\label{sec:ABC}

In this section, we review the basics of Lorentz-violating gravity, focusing on Einstein-\AE{}ther theory~\cite{Jacobson:2000xp} and khronometric gravity~\cite{Blas:2009qj}. We present only some basic information, and defer any details to the excellent reviews in~\cite{Jacobson:2004ts,Foster:2007gr,Yagi:2013ava, Mattingly:2005re, Jacobson:2008aj}. We here mostly follow the descriptions in~\cite{Yagi:2013ava}.

%--------------------------------------------------------------------------------
\subsection{The ABC of Einstein-\AE{}ther}

Einstein-\AE{}ther theory violates Lorentz symmetry gravitationally by introducing a vector field, $U^\mu$, of everywhere unit magnitude, that provides a ``preferred direction" to gravitation.  Up to quadratic terms in first derivatives of the field, the most generic action with such a field is~\cite{Yagi:2013ava}:
\begin{align}
\label{eq:EA-action}
S_{\AE{}} &= -\frac{1}{16\pi G_{\AE{}}}\int d^4x\sqrt{-g}(R + M^{\delta\sigma}_{\mu\nu}\nabla_\delta U^\mu\nabla_\sigma U^\nu) 
\nn \\
&+ S_{\rm mat}(\psi,g_{\mu \nu})\,,
\end{align}
where $S_{\rm mat}$ is the matter action, which depends on matter fields $\psi$ that couple directly to the metric $g_{\mu \nu}$. In the gravitational action, $G_{\AE{}}$ is the \emph{bare} gravitational constant in Einstein-\AE{}ther theory, $g$ is the determinant of the metric, $R$ is the Ricci scalar, and,
\begin{align}
	M^{\delta \sigma}_{\mu\nu} \equiv c_1g^{\delta \sigma}g_{\mu\nu}+c_2\delta^\delta_\mu\delta^\sigma_\nu+c_3\delta^\delta_\nu\delta^\sigma_\nu+c_4U^\delta U^\sigma g_{\mu\nu}\,,
\end{align}
with $\delta^{\mu}_\nu$ the Kronecker delta. 

Einstein-\AE{}ther theory is an interesting model, as it is the most general parity-preserving but Lorentz-violating theory that includes up to (quadratic) first derivatives in the vector field. Thus, a study of Einstein-\AE{}ther theory can be extended, via an appropriate choice of mappings, to many other, perhaps more restrictive, Lorentz-violating gravity theories.  Lorentz symmetry has, in some special cases, been shown to be a mechanism for renormalization of gravity~\cite{Visser:2009fg}, which makes these theories particularly interesting to study.

The strength of any Einstein-\AE{}ther modifications to GR depends on four coupling parameters $c_1$, $c_2$, $c_3$ and $c_4$. For convenience, we define the following (by now standard) combinations of these parameters:
\begin{align}
	c_{14} &\equiv c_1 + c_4\,, \qquad
	c_{123} \equiv c_1 + c_2 + c_3\,,
	\\
	c_+ &\equiv c_1 + c_3\,, \qquad c_- \equiv c_1 - c_3\,.
\end{align}
Other useful combinations of these parameters are
\begin{align}
	\label{eq:alpha1-EA}
	\alpha_1^{\ppN,\EA} &= -\frac{8(c_3^2+c_1c_4)}{2c_1-c_+c_-}\,,
	\\
	\label{eq:alpha2-EA}
	\alpha_2^{\ppN,\EA} &= \frac{\alpha_1^{\ppN,\EA}}{2}-\frac{(c_1+2c_3-c_4)(2c_1+3c_2+c_3+c_4)}{(2-c_{14})c_{123}}\,,
\end{align}
which control the magnitude of weak-field, preferred-frame effects in the parametrized post-Newtonian (ppN) formalism~\cite{will-living}. These quantities can be inverted to express the $c_{n}$'s as functions of $\alpha_{1,2}^{\ppN,\EA}$ and $c_{+,-}$, but the resulting expressions are long and unilluminating, so we will not present them explicitly here. Any linearly independent set of 4 of these quantities is enough to span the coupling parameter space of Einstein-\AE{}ther theory, but in this paper we will work with the set $(\alpha_{1}^{\ppN,\EA},\alpha_{2}^{\ppN,\EA},c_{+},c_{-})$.

Einstein-\AE{}ther theory passes all tests performed, provided its coupling parameters are small enough. Some of the most strict constraints come from Solar System observations and restrict the sizes of $\alpha_{1,2}^{\ppN,\EA}$~\cite{Jacobson:2008aj,will-living}. This, together with binary pulsar constraints~\cite{Yagi:2013qpa,Yagi:2013ava}, force  
\begin{align}
	|\alpha_1^{\ppN,\EA}| \lesssim 10^{-4}\,, \qquad |\alpha_2^{\ppN,\EA}| \lesssim 10^{-7}\,,
	\\
	c_+ \lesssim 10^{-2}\,, \qquad c_- \lesssim 10^{-3}\,.
\end{align}
Writing the $c_n$'s as functions of these quantities and saturating the bounds from the equations above, one finds
\begin{align}
	c_1 &\lesssim 10^{-2}%\frac{1}{2}c_++ \mathcal{O}(10^{-3})
	 \,, \qquad 
	c_2 \lesssim 10^{0}%-\frac{4c_+^2}{c_+^2-2\alpha_1^{\ppN,\EA}} +\mathcal{O}(10^{-1})\,,
	\\
	c_3 &\lesssim 10^{-2} %\frac{1}{2}c_+ %+\mathcal{O}(10^{-3})
	\,,
	\qquad
	c_4 \lesssim 10^{-2}\,. % \frac{1}{2}c_++\mathcal{O}(10^{-3})\,,
\end{align}

Many of the modifications to GR induced in Einstein-\AE{}ther theory depend on the \textit{sensitivities} of the bodies in question~\cite{Foster:2007gr}.  The sensitivity $s$ is a measure of the gravitational binding energy of a given body.  In the weak field limit, these sensitivities reduce to~\cite{Foster:2007gr,Yagi:2013ava},
\begin{align}
\label{eq:wf-s}
	s^{\EA} = -  \left(\alpha_1^{\ppN,\EA}-\frac{2}{3}\alpha_2^{\ppN,\EA}\right)\frac{C_*}{2}\,,
\end{align}
where $C_*$ is the compactness of the star. For the purposes of this paper, we use the numerically calculated sensitivities of~\cite{Yagi:2013ava}, which are valid for strongly self-gravitating objects. 

The conservative motion of massive objects is modified in Einstein-\AE{}ther theory from the GR prediction~\cite{Foster:2007gr}, as we will see explicitly in Sec.~\ref{sec:AE-theory}. The dominant effect is that the constant ${\cal{G}}$ that enters Kepler's third law in a binary system is not the same as the constant $G_{N}$ that enters Newton's third law in a Cavendish-type measurement, which in turn is different from the bare constant $G_{\EA}$ that enters the Einstein-\AE{}ther action~\cite{Foster:2007gr,Yagi:2013ava}. These constants are related, by
\begin{align}
\label{eq:Gs}
	G_{\EA} = G_N (1 - s_{1}) (1 - s_{2})\,, \qquad \mathcal{G} = \frac{2G_N}{2-c_{14}}\,,
\end{align}
where $s_{i}$ are the sensitivities. Since, by convention, we choose units in which $G_N = 1$, saturating the current bounds on the $c_{i}$'s, we see that $G_{\EA}-1 \leq 10^{-2}$ and $\mathcal{G}-1 \leq 10^{-4}$ for NSs.

Finally, in Einstein-\AE{}ther theory, the metric perturbation has five propagating degrees of freedom: two tensor modes, two vector modes, and one scalar mode~\cite{Jacobson:2004ts}. The propagation of the scalar and vector modes is responsible for dipole energy and angular momentum loss in binary systems. The propagation speeds of the scalar, vector and tensor modes in this theory is~\cite{Foster:2006az}
\begin{align}
\label{eqn:aespeed}
	w_0^{\EA} &= \left[\frac{(2-c_{14}c_{123})}{(2+3c_2+c_+)(1-c_+)c_{14}}\right]^{1/2}\,,
	\\
	w_1^{\EA} &= \left[\frac{2c_1-c_+c_-}{2(1-c_+)c_{14}}\right]^{1/2}\,,
	\\
	w_2^{\EA} &= \left(\frac{1}{1-c_+}\right)^{1/2}\,,
\end{align}
respectively. Note that the tensor propagation speed is different from the speed of light by a factor of $(1-c_+)^{-1/2}$, since we also work in units in which $c = 1$.  Note also that in order to avoid gravitational Cherenkov radiation and to enforce energy positivity, $w_0^{\EA}$, $w_1^{\EA}$ and $w_2^{\EA}$ must all be greater than or equal to one (see~\cite{Jacobson:2007fh} and references therein).

%--------------------------------------------------------------------------------
\subsection{The ABC of Khronometric Gravity}

Khronometric gravity is another Lorentz-violating theory, very similar to Einstein-\AE ther theory, except that in the former the Lorentz-violating vector field is required to be \emph{orthogonal} to hypersurfaces of a preferred time, defined through a foliation scalar $T$ (the ``khronon''). This orthogonality condition reduces the parameter space of the theory, which now only depends on three coupling parameters $(\lambda_{\KG},\beta_{\KG},\alpha_{\KG})$. The theory is then formally defined through the action
\begin{align}
S_{\KG} &= \frac{1-\beta_{\KG}}{16 \pi G_{\AE}} \int d^{3}x \; dT \; N \sqrt{h} \; \left(K_{ij} K^{ij} - \frac{1 + \lambda_{\KG}}{1 - \beta_{\KG}}  K^{2} 
\right. 
\nn \\
&+ \left.
\frac{1}{1 - \beta_{\KG}} {}^{(3)}R + \frac{\alpha_{\KG}}{1 - \beta_{\KG}} a_{i} a^{i}\right)
+ S_{\rm mat}(\psi,g_{\mu \nu})\,,
\end{align}
where again $S_{\rm mat}$ is the matter action, while $h$ is the determinant of the induced metric $h_{\mu \nu}$ on the hypersurfaces, $^{(3)}R$ is the Ricci scalar associated with this metric, $K_{ij}$ is the extrinsic curvature, and $a_{i} = \partial_{i} N$ is the acceleration of the lapse $N$, which is simply the preferred time component of the \AE ther co-vector field.

Khronometric gravity is interesting because it can be shown to be the low-energy limit of Horava gravity~\cite{Horava:2009uw}. The latter has been proposed as a quantum gravity model that is power-counting renormalizable~\cite{Visser:2009fg, Nishioka:2009iq}. In this paper, we focus only on the version of Horava gravity introduced in~\cite{Blas:2009qj,Blas:2009yd}, because here Minkowski spacetime is a ground-state and it reduces exactly to khronometric gravity at low energies. Einstein-\AE{}ther theory reduces to khronometric gravity in the limit as $c_- \to \infty$ (while keeping all other parameters finite) and with the mapping $\lambda_{\KG} = c_2$, $\beta_{\KG} = c_+$ and $\alpha_{\KG} = c_{14}$~\cite{Jacobson:2013xta}.

A useful combination of the coupling parameters of khronometric gravity is
\begin{align}
	\alpha_1^{\ppN,\KG} &= 4 \frac{\alpha_{\KG}-2\beta_{\KG}}{\beta_{\KG}-1}\,, 
	\\
	\alpha_2^{\ppN,\KG} &= \frac{\alpha_{\KG}-2\beta_{\KG}}{(\beta_{\KG}-1)(\lambda_{\KG}+\beta_{\KG})(\alpha_{\KG}-2)}
	\nn \\
	&\times
	[-\beta_{\KG}^2+\beta_{\KG}(\alpha_{\KG}-3)+\alpha_{\KG} 
	\nn \\
	&+ \lambda_{\KG}(-1-3\beta_{\KG}+2\alpha_{\KG})]\,,
\end{align}
which, as before, controls the magnitude of weak-field, preferred-frame effects in the ppN formalism~\cite{will-living}. These quantities can be inverted to express any two in the set $(\lambda_{\KG},\beta_{\KG},\alpha_{\KG})$ as functions of $\alpha_{1,2}^{\ppN,\KG}$. As in the Einstein-\AE{}ther case, any linearly independent set of 3 of these quantities is enough to span the coupling parameter space of khronometric gravity. In this paper, we will work with the set $(\alpha_{1}^{\ppN,\KG},\lambda_{\KG},\beta_{\KG})$, and we note that $\alpha_{\KG} = \alpha_{1}^{\ppN,\KG} (\beta_{\KG} - 1)/4 + 2 \beta_{\KG}$. Also note that $\alpha_{1,2}^{\ppN,\KG}$ both vanish when $\alpha_{\KG} = 2\beta_{\KG}$.

Khronometric gravity also passes all tests performed, provided its coupling parameters are small enough. Solar System~\cite{Jacobson:2008aj,will-living}, cosmological~\cite{Audren:2013dwa,Audren:2014hza} and binary pulsar observations~\cite{Yagi:2013ava} require that
\begin{align}
	|\alpha_1^{\ppN,\KG}| &\lesssim 10^{-4}\,, 
	\quad
	 \beta \lesssim 10^{-2}\,, \quad \lambda \lesssim 10^{-1}\,.
\end{align}
Saturating current constraints, this implies $\alpha_{\KG} \lesssim 10^{-2}$. 

Conservative orbital motion is modified in khronometric gravity in a similar way as in Einstein-\AE ther theory. In particular, the gravitational constant $G_{N}$ that enters into Cavendish-type measurements of Newton's third law is different from the bare constant $G_{\EA}$ that enters the khronometric gravity action and the constant ${\cal{G}}$ that enters Kepler's third law in binary systems. We will here express all such constants in terms of $G_{N}$ and then choose units via $G_{N} = 1$.

As in Einstein-\AE{}ther theory, dissipation during orbital motion is controlled by the evolution of all propagating degrees of freedom. The additional constraint of hypersurface orthogonality eliminates the vector modes, leaving only the scalar and tensor modes. The scalar mode is responsible for dipole energy and angular momentum loss.  The propagation speeds for the scalar and tensor modes are
\begin{align}
	w_{0}^{\KG} &= \left[\frac{(\alpha_{\KG}-2)(\beta_{\KG}+\lambda_{\KG})}{\alpha_{\KG}(\beta_{\KG}-1)(2+\beta_{\KG}+3\lambda_{\KG})}\right]^{1/2}\,,
	\\
	\label{eq:prop-speed-KG}
	w_2^{\KG} &= \left(\frac{1}{1-\beta_{\KG}}\right)^{1/2}\,,
\end{align}
respectively. Again, by stability considerations these speeds must be greater than or equal to one~\cite{Elliott:2005va}. 
%--------------------------------------------------------------------------------
\section{Response Functions in Modified Gravity Theories}
\label{sec:Response}

In this section, we review how to construct the Fourier transform of the response of an interferometer to a gravitational wave in generic modified gravity theories. We present this material for completeness and refer the interested reader to~\cite{bender,will-living,Yunes:2009yz,Chatziioannou:2012rf,PPE} and references therein for a more thorough coverage. 

In a generic modified gravity theory, the metric perturbation can possess up to 6 independent degrees of freedom: 2 tensor modes (spin-2), 2 vector modes (spin-1) and 2 scalar modes (spin-0). Given such a generic metric perturbation, the time-domain response function is (see e.g.~\cite{Chatziioannou:2012rf})
\begin{align}
	\label{eq:responsedecomp}
	h(t) = F_+h_++F_\times h_\times+F_\SE h_\SE+F_\SN h_\SN + F_\B h_\B+ F_\LL h_\LL\,.
\end{align}
where $(F_{+},F_{\times},F_{\SE},F_{\SN},F_{\B},F_{\LL})$ are beam pattern functions that can for example be found in~\cite{Chatziioannou:2012rf}. 

For gravitational waves emitted during the quasi-circular inspiral of compact objects, this response can be written as a sum over harmonics of the orbital phase $\Phi(t)$:
\begin{align}
	h(t) = \sum\limits_{\ell}^{}A_{\ell}(t)\left(e^{i \ell \Phi(t)} +e^{-i \ell \Phi(t)}\right)\,,
\end{align}
where $A_{\ell}(t)$ are some time-dependent amplitude coefficients that must be found by solving the modified field equations in the far-away, wave-zone.  For any $\ell$th harmonic, one generically has that $\frac{d}{dt}A_{\ell} \ll \frac{d}{dt}\left(e^{i \ell \Phi(t)} +e^{-i \ell \Phi(t)}\right)$ in the inspiral phase, where the orbital velocities are small. Moreover, for quasi-circular, non-spinning compact objects, the $\ell=2$ harmonic is dominant.

The evolution of the orbital phase is related to the binary's separation through the relativistic version of Kepler's third law. While not generically true, in Einstein-\AE{}ther theory and in khronometric gravity, Kepler's third law remains the familiar $P_{b}^2/r_{12}^3 = 4\pi^2/\mathcal{G}m$ to leading-order in a $v_{12} \ll 1$ expansion~\cite{Foster:2007gr}, where $P_{b}$ is the orbital period, $r_{12}$ is the orbital separation, $v_{12}$ is the magnitude of the relative orbital velocity vector, $m=m_{1}+m_{2}$ is the total mass of the binary and $\mathcal{G}$ is a two-body gravitational constant. Recall that ${\cal{G}}$ is corrected from $G_{N}$ via Eq.~\eqref{eq:Gs}. Thus, we can re-write the modified, relativistic version of Kepler's third law to 1PN order in the more convenient form 
\begin{align}
\label{eq:Kepler}
	r_{12} = \mathcal{G} \mathcal{M} \; \eta^{-1/5} \; u^{-2} \left[1 + \left(r_{12,1\PN}^{\GR} + r_{12,1\PN}^{\MG} \right) u^{2} + {\cal{O}}(c^{-3}) \right]\,,
\end{align} 
where $u \equiv \left(2\pi \mathcal{G} \mathcal{M}F\right)^{1/3} = {\cal{O}}(c^{-1})$ is a reduced (dimensionless) frequency, $F$ is the orbital frequency, ${\cal{M}} = m \; \eta^{3/5}$ is the chirp mass, $\eta = \mu/m$ is the symmetric mass ratio and $\mu = m_{1} m_{2}/m$ is the reduced mass. The terms $r_{12,1\PN}^{\GR}$ and $r_{12,1\PN}^{\MG}$ are the 1PN corrections to Kepler's law in GR and in Lorentz-violating gravity respectively.

Once the time-domain response has been found, one can obtain its frequency representation through its Fourier transform, defined here as follows:
\begin{align}
	\label{eq:spa}
	\tilde{h}(f) = \sum\limits_{\ell}^{}\int_{-\infty}^{+\infty} A_{\ell}(t) \left(e^{2\pi i f t+ i \ell \Phi(t)}+e^{2\pi i f t - i \ell \Phi(t)} \right) dt\,.
\end{align}
This generalized Fourier integral can be evaluated through the SPA, which uses the fact that the argument of the exponential varies much more rapidly than the amplitude, except in a small region in time-frequency that extremizes the phase~\cite{bender}.  The integral will thus be dominated by contributions around the stationary point $t_{0}$, defined via $2\pi i f - i \ell \dot{\Phi}(t_0)=0$. The first term in Eq.~\eqref{eq:spa} can be neglected by the Riemann-Lebesgue theorem~\cite{bender}, and the frequency-domain response in the SPA is simply
\begin{align}
	\label{eq:htwidle}
	\tilde{h}(f) = \sum\limits_{\ell}^{}\mathcal{A}_\ell \; e^{-i\Psi_{\ell}}\,,
\end{align}
where we have defined
\begin{align}
	\label{eq:amplitude}
	\mathcal{A}_{\ell} &\equiv \frac{A_{\ell}(t_0)}{\sqrt{\ell\dot{F}(t_0)}}\,,
	\qquad
	\Psi_\ell \equiv \int\limits_{}^{F(t_0)} \nu(F') \; dF' +\frac{\pi}{4}\,, 
\end{align}
$F(t) = \dot{\Phi}(t)/(2 \pi)$ is the orbital frequency, overhead dots stand for time differentiation, and the integrand
\begin{align}
	\label{eq:D(F)}
	\nu(F) &\equiv 2\pi\left(\ell\frac{F}{\dot{F}(F)}-\frac{f}{\dot{F}(F)}\right)\,.
\end{align}
See for example~\cite{Yunes:2009yz} for more details. 

Clearly, in order to compute the frequency-domain response function in the SPA, we must first find the rate of change of the orbital frequency: $\dot{F}$.  We can do this via the chain rule
\begin{align}
	\label{eq:Fdot}
	\dot{F} &= \frac{dF}{dE_{b}} \frac{dE_{b}}{dt}\,,
\end{align}
where $E_{b}$ is the binary's binding energy. Assuming a modified gravity theory that is semi-conservative~\cite{will-living}, a balance law must then exist to relate the amount of energy the binary system loses per unit time, $\dot{E}_{b}$, to the amount of energy carried away by all propagating degrees of freedom ${\cal{L}}$, i.e.~$\dot{E}_{b} = -{\cal{L}}$.

Now, in order to proceed with the calculation, we will need the binding energy, $E$ and the rate of change of the binding energy, $\dot{E}_{b}$.  Expanding in powers of $v_{12}/c$, we can write these quantities as
\begin{align}
	\label{eqn:edotgen}
	\dot{E}_{b} &= \dot{E}^{\GR}_{0\PN} \left[ \dot{E}^{\MG}_{-1\PN} v^{-2} + 1 + \dot{E}^{\MG}_{0\PN}  +\mathcal{O}(c^{-2})\right]\,,
	\\
	E_b &= E_{0\PN}^{\GR} \left[1 + \left( E_{1\PN}^{\GR} +  E^{\MG}_{1\PN}\right) v^2 + \mathcal{O}(c^{-4})\right] \,,
\end{align}
where $E^{\GR}_{0\PN}$ and $\dot{E}^{\GR}_{0\PN}$ are the expected values in pure GR at 0PN order, $\dot{E}^{\MG}_{-1\PN}$ is a modified gravity correction induced by dipole radiation, and $\dot{E}^{\MG}_{0\PN}$ is a correction to quadrupole radiation and a 1PN correction to dipole radiation. Similarly, $E^{\MG}_{1\PN}$  is a modified gravity correction to the binding energy at 1PN order.  We will examine how all of these terms specifically affect the dynamics of the system in Lorentz-violating gravity theories in the following section.

Let us now assume that the modified theory of gravity is a small deformation from GR, i.e.~that we can expand all functions as the GR expectation plus a small deformation. Let us further assume that the deformation depends on certain coupling parameters continuously, such that in the limit as these coupling parameters vanish, the deformation also vanishes. With this at hand and using the modified and relativistic version of Kepler's law, we can then always expand $\dot{E}_{b}$ and $E_{b}$ as
\begin{align}
\label{eq:Edot-def}
	\dot{E}_{b}(u) &= \dot{E}_{b}^{\GR}(u) \left[1 + \delta_{\dot{E}_{b}}(u) \right]\,,
	\\
\label{eq:Eb-def}
	{E}_{b}(u) &= {E}_{b}^{\GR}(u) \left[1 + \delta_{{E}_{b}}(u)\right] \,,
\end{align}
where $\dot{E}^{\GR}_{b}(u) \equiv - (32/5) u^{10} [1 + {\cal{O}}(c^{-2})]$ and ${E}^{\GR}_{b}(u) \equiv -(1/2) {\cal{M}} u^{2} [1 + {\cal{O}}(c^{-2})]$ are the GR prediction to as high a PN order as one wishes to keep (see e.g.~\cite{Blanchet:2002av}), while $\delta_{\dot{E}_{b}}(u)$ and $\delta_{{E}_{b}}(u)$  are small deformations. Putting these two equations together, we then find
\begin{align}
	\label{eq:Fdot-def}
	\dot{F}(u) &= \dot{F}_\GR(u) \left[1 + \delta_{\dot{F}}(u)\right] \,,
	\\
	\nu(u) &= \nu_{\GR}(u) \left[1 + \delta_\nu(u)\right]\,,
\end{align}
where $\dot{F}_{\GR}(u) \equiv 48/(5\pi) u^{11}/\mathcal{M}^2 [1 + {\cal{O}}(c^{-2})]$ and $\nu_{\GR}(u) = (5\pi^2/24) \mathcal{M}^2 u^{-11} (F\ell-f) [1 + {\cal{O}}(c^{-2})]$  are the GR predictions, and again $\delta_{\dot{F}}(u)$ and $\delta_{\nu}(u)$ are small deformations. In separating the GR prediction from its deformation, one must be careful to account for the deformation that is generated due to correction to Kepler's law [Eq.~\eqref{eq:Kepler}], since ${\cal{G}} \neq G_{N}$ in general. 

For future convenience, let us introduce the parametrized post-Einsteinian (ppE) framework~\cite{PPE}. This is a scheme for modeling many different modified gravity theories, in close analogy to the ppN framework. This parametrization relies on the fact that, while in theory the set of possible modifications to GR is uncountably infinite, we wish to focus on those that are (i) well-motivated from fundamental physics, (ii) pass all weak-field, Solar System tests and (iii) lead to non-negligible effects in the non-linear and dynamical regime, e.g.~during the late inspiral and merger of binary systems. The ppE framework is able to reproduce gravitational wave predictions for all known, well-motivated modified gravity theories that can be treated as small deformations from the GR prediction.  

The simplest ppE waveform model for the quasi-circular inspiral of compact objects is~\cite{PPE}
\be
\tilde{h}_{\ppE}(f) = \sum_{\ell=1}^{\infty} {\cal{A}}^{(\ell)}_{\ppE}(f) \; e^{i \Psi^{(\ell)}_{\ppE}(f)}\,,
\ee
with the $\ell$-harmonic, ppE phase and amplitude~\cite{PPE}
\begin{align}
	\label{eq:appE}
		{\cal{A}}^{(\ell)}_{\ppE}(f) &= {\cal{A}}^{(\ell)}_{\GR,\PN}(f) \left[1+ u_{\ell}^{a_{\ppE}}  \sum_{k=0}^{\infty} \alpha_{\ppE,k} u_{\ell}^{k} \right]\,,
	\\
	\label{eq:psippE}
	\Psi^{(\ell)}_{\ppE}(f) &= \Psi^{(\ell)}_{\GR,\PN}(f) + u_{\ell}^{b_{\ppE}} \sum_{k=0}^{\infty} \beta_{\ppE,k} u_{\ell}^{k}\,,
\end{align}
and where ${\cal{A}}^{(\ell)}_{\GR,\PN}(f)$ and $\Psi^{(\ell)}_{\GR,\PN}(f)$ are the $\ell$-harmonic of the GR amplitude and phase prediction in the PN approximation for an inspiral waveform, 
$(b_{\ppE},a_{\ppE})$ are leading-order, ppE exponent parameters,
$(\alpha_{\ppE,0},\beta_{\ppE,0})$ are leading-order, ppE amplitude parameters, 
and $(\alpha_{\ppE,k\geq1},\beta_{\ppE,\geq1})$ are higher PN order, ppE amplitude parameters.
Note that for non-spinning and quasi-circular inspirals, $\ell=2$ is the dominant harmonic in GR, 
but this need not be the case in modified gravity theories. Of course, GR waveforms are obtained when all ppE amplitude parameters $\alpha_{\ppE,k}$ and $\beta_{\ppE,k}$ are simultaneously zero. 

An interesting simplification that is usually made in the literature is that of the \emph{restricted PN} approximation. This approximation requires that we keep only the leading PN order terms in the amplitude, but all available PN corrections to the phase (see e.g.~\cite{Buonanno:2009zt}). When the $\ell=2$ harmonic is dominant, one then has the fully-restricted PPE waveform as
\be
\tilde{h}_{\fr,\ppE}(f) = {\cal{A}}^{(2)}_{\ppE,\Newt}(f) \; e^{i \Psi^{(2)}_{\ppE}(f)}\,,
\ee
where ${\cal{A}}^{(2)}_{\ppE,\Newt}(f)$ is given by Eq.~\eqref{eq:appE} with $\ell=2$ and keeping only the $k=0$ term in the sum, while $\Psi^{(2)}_{\ppE}$ is given by Eq.~\eqref{eq:psippE} with $\ell=2$ and keeping as many PN terms as possible. More complicated ppE models have been proposed that incorporate the merger and ringdown phases, but in this paper we will focus solely on the inspiral part. A further discussion of the ppE framework can be found in~\cite{PPE,Chatziioannou:2012rf, Loutrel:2014vja}, while data analysis implementations have been worked out in~\cite{cornishsampson, Yunes:2010qb, Li:2011cg,Agathos:2013upa, Sampson:2013lpa,Sampson:2013jpa,Sampson:2013wia,Huwyler:2014vva}.

%---------------------------------------------------------------------------------
\section{Response Functions in Lorentz Violating Gravity: Einstein-\AE{}ther Theory}
\label{sec:AE-theory}

In this section, we construct the time-domain and frequency-domain representations of the response function in Einstein-\AE{}ther theory. We use heavily the generic analysis presented in Sec.~\ref{sec:Response} and assume the coupling parameters $(\alpha_1^{\ppN, \EA},\alpha_2^{\ppN, \EA},c_{+},c_{-})$ are all smaller than unity, as required by current constraints. Given the work of~\cite{Yagi:2013qpa,Yagi:2013ava}, this also implies that the sensitivities $s_{i} \ll 1$ and, for NS binaries in particular, $(s_1-s_2) \ll 1$.

\subsection{Time Domain Response}

In Einstein-\AE{}ther theory, the metric perturbation has only five propagating degrees of freedom: two tensor modes, two vector modes, and one scalar mode~\cite{Jacobson:2004ts}. These degrees of freedom are encoded in the metric perturbation, which can be written in harmonic coordinates and in an appropriate gauge as~\cite{Foster:2007gr}
\begin{align}
	h^{\EA}_{ij}&=\phi_{ij}^{\EA}+\textrm{TT}_{ij}(\mathcal{F}_{\EA}) +\phi_{,ij}^{\EA}\,,
	\\
	h^{\EA}_{0i}&=\gamma_i^{\EA}\,,
	\\
	h^{\EA}_{00}&=\frac{1}{c_{14}}\mathcal{F}_{\EA}\,,
\end{align}
with the gauge condition $\phi^{\EA}_{ij,j} = \phi^{\EA}_{ii} = \gamma^{\EA}_{i,i} =0$. Here $\textrm{TT}_{ij}$ is the transverse-traceless operator, and, to leading PN order~\cite{Foster:2007gr} 
\begin{align}
	\phi^{\EA}_{ij}&=4  G_{\EA} \frac{\mu}{ r}\left(v_{12}^iv_{12}^j- \frac{m}{r_{12}} n_{12}^in_{12}^j\right)\,,
	\\
	\label{eqn:fcurly}
	\mathcal{F}_{\EA} &= -8  G_{\EA} \frac{\mu}{r} \frac{(s_1^{\EA}-s_2^{\EA})}{w_0^{\EA}(c_{14}-2)}n_{i} v_{12}^i\,,
	\\
	\gamma^{i}_{\EA}&=  4 G_{\EA} \frac{\mu}{r} \frac{c_{+}(s_1^{\EA}-s_2^{\EA})}{2c_1-c_{-}c_{+}}\left(n_jv_{12}^j n^{i}+v_{12}^i\right)\,.
	\\
	\phi^{\EA}_{,ij} &= - \frac{1 + c_{2}}{c_{123}} {\cal{F}}_{\EA}\,,
	\label{eq:phi}
\end{align}
where $r_{12}^i$ is the relative position vector, $n_{12}^{i}$ is its unit vector, $r$ is the distance from the center of mass to the observer, $n^{i}$ is its unit vector, and the relative orbital velocity of the binary is $v_{12}^i$. Recall that the quantities $s_{1}^{\EA}$ and $s_{2}^{\EA}$ are the sensitivities of the two stars.

Using this along with~\cite{Chatziioannou:2012rf}, the polarizations are
\begin{align}
	h_b^{\EA}&=4 \; G_{\EA} \frac{(s_1-s_2)}{w_0^{\EA}(c_{14}-2)}\left[\frac{1}{c_+}+\frac{(c_2+1)}{c_{123}}\right] 
	\\
	&\times   \sin\iota \; \eta^{1/5} \; \frac{\mathcal{M}}{r} u \; \cos(\Phi) \,,
	\nonumber
	\\
	h^{\EA}_\ell &= -8 \; G_{\EA} (s_1-s_2)\left[2+  \frac{c_{123} + c_{+} (c_2+1)}{w_0^{\EA}(c_{14}-2) c_{+} c_{123}}\right]
	\\
	&\times \sin\iota \; \eta^{1/5} \; \frac{\mathcal{M}}{r} u \; \cos(\Phi)
	\,,
	\nonumber
	\\
	h^{\EA}_{sn}&=4 \; G_{\EA} \frac{c_+ (s_1-s_2)}{c_+c_- - 2c_1} \; \eta^{1/5} \; \frac{\mathcal{M}}{r} u \; \sin(\Phi)\,,
	\\
	h^{\EA}_{se}&=4 \; G_{\EA} \frac{c_+ (s_1-s_2)}{c_+c_- - 2c_1} \; \cos\iota \; \eta^{1/5} \; \frac{\mathcal{M}}{r} u \;  \cos(\Phi)\,,
	\\
	h_{+}^{\EA} &= 2 \; G_{\EA} (1+\cos^2\iota) \; \eta^{1/5} \; \frac{\mathcal{M}}{r} u^2 \;  \cos(2\Phi)\,,
	\\
	h_{\times}^{\EA} &= 4  G_{\EA} \cos\iota  \; \eta^{1/5} \; \frac{\mathcal{M}}{r} u^2 \; \sin(2\Phi)\,,
\end{align}
to leading PN order, where $\iota$ is the inclination angle and recall $\Phi$ is the orbital phase. Notice that the amplitude of non-GR terms is ${\cal{O}}(v_{12})$ larger than the amplitude of GR terms. This is a common feature of vector and scalar modes in modified gravity theories~\cite{Chatziioannou:2012rf}. Note also that $h_b^{\EA}$ and $h_l^{\EA}$ are functions only of $\cos \Phi$ (the first-harmonic of the orbital phase), and that they have the same exact time-dependence, so that their ratio is actually time-independent.

The response function in the time-domain can then be written as
\begin{align}
	\label{eq:responset}
	h_{\EA}(t) &= A^{\EA}_2 \frac{\mathcal{M}}{r}u^2\left(e^{-2i\Phi+i\Theta}+ e^{2i\Phi-i\Theta} \right) \\
	&+ A^{\EA}_1 \; \bar{\alpha}^{\EA} \;  \frac{\mathcal{M}}{r}\eta^{1/5}u\left(e^{-i\Phi}+ e^{+i\Phi} \right)\,,
	\nonumber
\end{align}
where $\Theta = \tan^{-1}[2F_{\times}\cos\iota/F_+(1+\cos^2\iota)]$, and to leading PN order, 
\begin{align}
\label{eq:A2}
	A^{\EA}_2 &= G_{\EA}\left[F^2_+(1+\cos^2\iota)^2+4F^2_{\times}\cos^2\iota\right]^{1/2} \,,
	\\
	A_1^{\EA} &= 2G_{\EA}(s_1-s_2)\,,
	\\	
	\bar{\alpha}^{\EA} &\equiv \left[ \frac{4c_+}{2c_1-c_+c_-}F_\ell - \frac{c_2+1}{c_{123}(c_{14}-2)w_0^{\EA}}\right.
		\nn \\
		&\left.\times \left(F_b+2F_\ell\right)
		+\frac{1}{(c_{14}-2)c_+w_0^{\EA}}\left(F_b-2F_\ell\right)\right]\sin\iota
		\nonumber
		\\
		& + \frac{c_+}{2c_1-c_+c_-}\left[iF_{\SN}+\cos\iota F_{\SE}\right]\,.
			\label{eq:alpha}
\end{align}
Note that the term in the first line of Eq.~\eqref{eq:responset}, proportional to the second harmonic of the orbital phase, is precisely that expected in GR in the limit $G_{\EA} \to 1$~\cite{Chatziioannou:2012rf}. On the other hand, the term in the second line of Eq.~\eqref{eq:responset}, proportional to the first harmonic of the orbital phase, goes to zero in the GR limit because then the sensitivities go to zero. Finally, note that the $\ell=2$ amplitude is of $\mathcal{O}(c^{-1})$ smaller than the $\ell=1$ term, but the latter is multiplied by a function of the coupling parameters, so it actually is parametrically smaller.

%---------------------------------------------------------------------------------
\subsection{Frequency Domain Response}

Let us begin with the calculation of the Fourier phase.
As discussed in Sec.~\ref{sec:Response}, this quantity in the SPA is controlled by the rate of change of the orbital frequency, which in turn depends on the rate of change of the binding energy. Using Eq.~\eqref{eq:Edot-def}, we write the correction to the rate of change of the gravitational binding energy as~\cite{Yagi:2013ava} 
\begin{align}
	\label{eq:Edot}
	\delta_{\dot{E}_{b}}^{\EA} &= \frac{7}{4}\eta^{2/5} \dot{E}^{\EA}_{-1\PN} u^{-2} + \dot{E}^{\EA}_{0\PN} \,,
\end{align}
where $\dot{E}^{\EA}_{-1\PN}$ and $\dot{E}^{\EA}_{0\PN}$ are functions of the $c_{i}$'s of Einstein-\AE{}ther theory that we determine next.

For the systems examined here, i.e.~NS binaries with $C_* < 1/2$, in the small coupling limit, the sensitivities can be shown to be small~\cite{Yagi:2013ava}.  In the regions examined in this paper, the sensitivities are generally less than $\mathcal{O}(10^{-4})$ each, so that $(s_1-s_2)^2 \lessapprox 10^{-8}$, making this a small parameter indeed.  As such we will ignore terms proportional to powers of $(s_1-s_2)^n$ for $n>2$. Then, for Einstein-\AE{}ther theory, we find that
\begin{align}
	\label{eqn:bm1pn}
	\dot{E}^{\EA}_{-1\PN} \equiv \tilde{\beta}_{-1\PN}^{\EA} &= \frac{5}{84}\mathcal{G}(s_1-s_2)^2\frac{(c_{14}-2)(w_0^{\EA})^3-(w_1^{\EA})^3}{c_{14}(w_0^{\EA})^3 (w_1^{\EA})^3}\,,
	\\
	\label{eqn:b0pn}
	\dot{E}^{\EA}_{0\PN} \equiv \tilde{\beta}_{0\PN}^{\EA} &= \mathcal{G}\left(1-\frac{c_{14}}{2}\right)\left(\frac{1}{w_2^{\EA}}+\frac{2c_{14}c_+^2}{(2c_1-c_-c_+)^2 w_1^{\EA}}\right.
	\nonumber
	\\
	&\left. +\frac{3c_{14}(Z_{\EA}-1)^2}{2w_0^{\EA}(2-c_{14})}+S\mathcal{A}_2+S^2\mathcal{A}_3\right) - 1\,,
\end{align}
where $S\equiv (s_1 m_2 + s_2 m_1)/m$, while $\mathcal{A}_2$ and $\mathcal{A}_3$ are functions of the coupling parameters, given explicitly in~\cite{Yagi:2013ava}, and
\begin{align}
\label{eq:ZEA}
	Z_{\EA}&=\frac{(\alpha_1^{\ppN}-2\alpha_2^{\ppN})(1-c_+)}{3(2c_+-c14)}\,.
\end{align}
In Sec.~\ref{sec:Fisher}, it will become clear why we have redefined $\dot{E}^{\EA}_{-1,0\PN}$ as $\tilde{\beta}^{\EA}_{-1,0\PN}$. 

The binding energy, $E_{b}$, was also presented explicitly in~\cite{Yagi:2013ava}, and it is not modified to leading PN order from the GR result. The 1PN modification to the binding energy has not yet been calculated and it would depend not only on $s_{A}$ but also on its derivative with respect to the scalar field, $s_{A}'$~\cite{Foster:2007gr}. The latter is a function of $s_{A}$ but also of the second derivative of the binary's masses with respect to the scalar field, which has not yet been calculated. 

Our ignorance of the 1PN corrections to the binding energy, however, should not affect the validity of the results presented here.  These corrections will enter our calculation proportional to both $u^{2} \times (s_1-s_2)^2$. For systems where the difference in sensitivities is very small, such as NS binaries, these terms will be strictly sub-dominant relative to other 1PN terms that enter through the rate of change of the binding energy, in our small-coupling expansion. For example, the $S^{2} \mathcal{A}_3$ term is much larger than terms proportional to $(s_{1} - s_{2})^{2} u^{2}$, since $S^{2} \gg (s_{1} - s_{2})^{2}$ for systems with similar sensitivities. For this reason, we will here ignore $r_{12,1\PN}^{\MG}$ and $\delta _{E_{b}}$ in Eqs.~\eqref{eq:Kepler} and~\eqref{eq:Eb-def}, but such modifications can easily be included in the future, when $s_{A}'$ is calculated.  

Combining all of this, we can not compute the Fourier response function in the SPA from Eq.~\eqref{eq:htwidle}. Using Eqs.~\eqref{eq:Fdot-def} and~\eqref{eq:D(F)}, the correction to the frequency evolution is
\begin{align}
	\delta_{\dot{F}}^{\EA} &=  \frac{7}{4} \eta^{2/5}\tilde{\beta}^{\EA}_{-1\PN} u^{-2} + \tilde{\beta}^{\EA}_{0\PN}\,.
\end{align}
while $\nu(F)$ is
\begin{align}
	\delta_\nu^{\EA} &= \left(-\frac{5}{48} \eta^{2/5}\tilde{\beta}_{-1\PN}^{\EA}u^{-2} + \tilde{\beta}^{\EA}_{0\PN} \right) (F\ell-f)\,.
\end{align}
With this, the SPA Fourier phase is
\begin{align}
	\Psi^{\EA}_{\ell} &= 2\pi ft_c+\Phi_c-\frac{\pi}{4}  -\frac{3\ell}{256}u_{\ell}^{-5} \left[1 + {\cal{O}}(c^{-2}) \right]
	\nonumber
	\\
	\label{eqn:eaphase}
	&-\frac{3\ell}{256}u_{\ell}^{-5} \left[\tilde{\beta}_{-1\PN}^{\EA}\eta^{2/5}u_{\ell}^{-2} +  \tilde{\beta}_{0\PN}^{\EA} + {\cal{O}}(c^{-2}) \right]\,,
\end{align}
where $(t_{c},\phi_{c})$ are constant time and phase offsets (sometimes called the time and phase of coalescence) and where we have defined $u_\ell \equiv \left(2\pi\mathcal{G} \mathcal{M}f/\ell\right)^{1/3}$. Notice that $u_{\ell} \neq u$,  where the former is defined in terms of the gravitational wave frequency $f$, while the latter depends on the orbital frequency $F$. 

We can now make several observations on the important result presented in Eq.~\eqref{eqn:eaphase}. First, the first line in this equation is the GR prediction for the SPA phase, while the second line is the Einstein-\AE{}ther correction. Notice that the latter depends on the quantities $\tilde{\beta}_{-1,0\PN}^{\EA}$, which are functions of the coupling parameters defined precisely so that Eq.~\eqref{eqn:eaphase} takes the simple form presented above. Second, the gravitational wave phase reduces to the GR prediction as $\tilde{\beta}_{-1,0\PN}^{\EA} \to 0$. This implies that a gravitational wave detection consistent with GR would allow us to constrain the particular combination of coupling parameters in $\tilde{\beta}_{-1,0\PN}^{\EA}$. Third, the GR phase was truncated to leading order with ${\cal{O}}(c^{-2})$ remainders to simplify the presentation. One can easily include as many higher PN order terms to the GR phase as one wishes, for example using~\cite{Buonanno:2009zt}. The Einstein-\AE{}ther correction to the SPA phase was truncated at ${\cal{O}}(c^{-4})$ relative to its leading-order term, but it is also proportional to $\tilde{\beta}_{-1,0\PN}^{\EA}$, and thus, parametrically smaller than GR terms of the same PN order.

Similarly, we can compute the amplitude of the Fourier response in the SPA:
and
\begin{align}
	\mathcal{A}_1^{\EA} &= -\left(\frac{5\pi}{48}\right)^{1/2}  A^{\EA}_1\bar{\alpha}^{\EA} \frac{\mathcal{M}^2}{r_{12}}\eta^{1/5} u_1^{-9/2}\,,
	\\
	\mathcal{A}_2^{\EA} &= \left(\frac{5\pi}{96}\right)^{1/2} A^{\EA}_2\frac{\mathcal{M}^2}{r_{12}}u_2^{-7/2}\,.
\end{align}
Notice again that ${\cal{A}}_{1}^{\EA}$ vanishes and ${\cal{A}}_{2}^{\EA}$ reduces to the GR prediction as the coupling parameters go to zero. Note also that the amplitude of the $\ell=1$ harmonic, ${\cal{A}}_{1}^{\EA}$ is larger than that of the $\ell=2$ harmonic by a factor of ${\cal{O}}(v/c)$, but parametrically smaller due to its dependence on $\bar{\alpha}^{\EA}$. 

Let us now quickly review which terms have been accounted for and which are left as remainders.  Recall, to -1PN order, our expression is complete.  However, to 0PN order, we expand in powers of $(s_1-s_2)^2$.  Thus these final expressions we have obtained are valid only for systems where the sensitivity difference is small. Thus, the calculation would have to be redone from Eq.~\eqref{eq:responset} if, for example one considers mixed binaries and the individual sensitivities are large. This, however, would require computation of the derivative of the sensitivities, which is currently unknown.

For convenience, we can also rewrite the Fourier transform as in Eq.~\eqref{eq:responsedecomp}:
\begin{align}
	\label{eqn:responsedecompea}
	\tilde{h}(f)^{\EA}_{l} &=\sin\iota\left(\frac{5\pi}{48}\right)^{1/2} G_{\EA}\alpha^{\EA}_l\frac{\mathcal{M}^2}{r_{12}}\eta^{1/5}u_1^{-9/2} e^{i\Psi_1}\,,
	\\
	\tilde{h}(f)^{\EA}_{b} &=\sin\iota\left(\frac{5\pi}{48}\right)^{1/2} G_{\EA}\alpha^{\EA}_b\frac{\mathcal{M}^2}{r_{12}}\eta^{1/5}u_1^{-9/2} e^{i\Psi_1}\,,
	\\
	\tilde{h}(f)^{\EA}_{\SE} &=\cos\iota\left(\frac{5\pi}{48}\right)^{1/2} G_{\EA}\alpha^{\EA}_v\frac{\mathcal{M}^2}{r_{12}}\eta^{1/5}u_1^{-9/2} e^{i\Psi_1}\,,
	\\
	\tilde{h}(f)^{\EA}_{\SN} &=\left(\frac{5\pi}{48}\right)^{1/2} G_{\EA}\alpha^{\EA}_v\frac{\mathcal{M}^2}{r_{12}}\eta^{1/5}u_1^{-9/2}e^{i(\Psi_1+\pi/2)}\,,
	\\
	\tilde{h}^{\EA}_+(f) &= (1+\cos^2 \iota)\left(\frac{5\pi}{96}\right)^{1/2}G_{\EA}\frac{\mathcal{M}^2}{r_{12}}u_2^{-7/2}e^{i\Psi_2}\,,
	\\
	\label{eq:lastEA}
	\tilde{h}^{\EA}_\times(f) &=2\cos(\iota)  \left(\frac{5\pi}{96}\right)^{1/2}G_{\EA}\frac{\mathcal{M}^2}{r_{12}}u_2^{-7/2}e^{i(\Psi_2+\pi/2)}\,.
\end{align}
where
\begin{align}
	\alpha^{\EA}_v &\equiv 2G_{\EA}(s_1-s_2)\frac{c_+}{2c_1-c_+c_-}\,,
	\\
	\alpha^{\EA}_b &\equiv 2G_{\EA}(s_1-s_2)\frac{1}{(c_{14}-2)w_0^{\EA}}\left[\frac{1}{c_+}-\frac{c_2+1}{c_{123}}\right]\,,
	\\
	\alpha^{\EA}_l &\equiv -4G_{\EA}(s_1-s_2)\frac{1}{(c_{14}-2)w_0^{\EA}}\left[\frac{1}{c_+}+\frac{c_2+1}{c_{123}}\right]
	\nonumber
	\\
	 &+4\alpha^{\EA}_v\,.
\end{align}

Comparing the Einstein-\AE{}ther Fourier phase and amplitude to the ppE Fourier amplitude and phase in Eqs.~\eqref{eq:appE} and~\eqref{eq:psippE}, we find the mapping to a fully-restricted ppE waveform:
\begin{align}
	\label{eq:ppE}
	\alpha_{\ppE,0} &= G_{\EA}-1\,, \qquad
	a_{\ppE} = 0\,,
	\\
	\beta_{\ppE, 0} &= - \frac{3}{128} \tilde{\beta}^{\EA}_{-1\PN}\eta^{2/5}\,, \qquad
	b_{\ppE} = -7\,,
	\\
	\beta_{\ppE, 1} &= 0\,, \qquad\qquad	\beta_{\ppE,2} = - \frac{3}{128} \tilde{\beta}^{\EA}_{0\PN}\,.
\end{align} 
Note that $\alpha_{\ppE,0} \ll 1$ and $\beta_{\ppE,0,1} \ll 1$ as expected. Note also that the leading PN order correction to the ppE phase enters at $-1$PN order, as expected. 

%\begin{align}
%	\label{eq:appE}
%		A^{(2)}_{\ppE}(f) &= A^{(2)}_{\GR,\PN} \left[1+ u_{2}^{a_{\ppE}}  \alpha_{\ppE,0} \right]\,,
%	\\
%	\label{eq:psippE}
%	\Psi^{(2)}_{\ppE}(f) &= \Psi^{(2)}_{\GR,\PN} + u_{2}^{b_{\ppE}} (\beta_{\ppE,0} + \beta_{\ppE,1} u_{2} + \beta_{\ppE,2} u_{2}^{2})\,,
%\end{align}
%
%\begin{align}
%	\Psi^{\EA}_{\ell} &= 2\pi ft_c+\Phi_c-\frac{\pi}{4}  -\frac{3\ell}{256}u_{\ell}^{-5} \left[1 + {\cal{O}}(c^{-2}) \right]
%	\nonumber
%	\\
%	&-\frac{3\ell}{256}u_{\ell}^{-5} \left[\tilde{\beta}_{-1\PN}^{\EA}\eta^{2/5}u_{\ell}^{-2} +  \tilde{\beta}_{0\PN}^{\EA} + {\cal{O}}(c^{-2}) \right]\,,
%\end{align}
%
%
%---------------------------------------------------------------------------------
\section{Response Functions in Lorentz Violating Gravity: Khronometric Gravity}
\label{sec:KG-theory}

This section parallels Sec.~\ref{sec:AE-theory}, but we focus here on khronometric gravity instead of Einstein-\AE{}ther theory. As before, we rely heavily on Sec.~\ref{sec:Response} for the construction of the Fourier transform of the gravitational wave response in the SPA.

%---------------------------------------------------------------------------------
\subsection{Time Domain Response}

In khronometric gravity, the two vector degrees of freedom vanish and the metric perturbation has now only four polarizations, 
two less than in Einstein-\AE{}ther theory due to the hypersurface-orthogonality constraint. 
In harmonic coordinates and in a suitable gauge, the metric has components
\begin{align}
	h_{ij}^{\KG}&=\phi^{\KG}_{ij}+\textrm{TT}_{ij}(\mathcal{F}_{\KG})  +\phi_{,ij}^{\KG} \,,
	\\
	h_{00}^{\KG}&=\frac{1}{\alpha_{\KG}}\mathcal{F}_{\KG}\,,
\end{align}
with the gauge condition $\phi^{\KG}_{ij,j} = \phi^{\KG}_{ii} = \gamma^{\KG}_{i,i} =0$.
From Eq.~(119) in~\cite{Blas:2011zd,Yagi:2013ava}, to leading PN order we have
\begin{align}
	&\phi^{ij}_{\KG} = 4G_{\AE}\frac{\mu}{r}\left(v_{12}^i v_{12}^j-\frac{m}{r_{12}}n_{12}^in_{12}^j\right)\,,
	\\
	& \mathcal{F}_{\KG} = -8G_{\AE}\frac{\mu}{r}\frac{(s_1^{\KG}-s_2^{\KG})}{w_{0}^{\KG}(\alpha_{\KG}-2)}n_{i} v_{12}^i\,,
	\\
	& \phi_{,ij}^{\KG} = -\frac{1+\lambda_{\KG}}{\beta_{\KG}+\lambda_{\KG}}\mathcal{F}_{\KG}\,,
\end{align}
where $(\mu,m,r,n^{i},r_{12},v_{12}^{i},n_{12}^{i})$ are defined as in Einstein-\AE{}ther theory (see definitions below Eq.~\eqref{eq:phi}), while $s_{1}^{\KG}$ and $s_{2}^{\KG}$ are sensitivities in khronometric gravity, which have been computed numerically in~\cite{Yagi:2013qpa,Yagi:2013ava} for NSs.

Following the convention of~\cite{Chatziioannou:2012rf}, we find that the gravitational wave polarization modes are
\begin{align}
	h_b^{\KG} &= 4 \; G_{\AE} \;
	\frac{(s_1^{\KG}-s_2^{\KG})}{\alpha^{\KG}-2}\sqrt{\frac{(\beta^{\KG}-1)(2+\beta^{\KG}+3\lambda^{\KG})}{(\alpha^{\KG}-2)(\beta^{\KG}+\lambda^{\KG})}}
	\nonumber
	\\
	&\times
	\left[\frac{1}{\sqrt{\alpha^{\KG}}}+\frac{\sqrt{\alpha^{\KG}}(\lambda^{\KG}+1)}{(\beta^{\KG}+\lambda^{\KG})}\right]
	\sin\iota \; 
	\eta^{1/5} \; \frac{\mathcal{M}}{r} \; u \; \cos\Phi\,,
	\\
	h_l^{\KG} &= - 8 \; G_{\AE} \; 
	\frac{(s_1^{\KG}-s_2^{\KG})}{\alpha^{\KG}-2} \sqrt{\frac{(\beta^{\KG}-1)(2+\beta^{\KG}+3\lambda^{\KG})}{(\alpha^{\KG}-2)(\beta^{\KG}+\lambda^{\KG})}} \; 
	\nonumber \\
	&\times \left[\frac{1}{\sqrt{\alpha^{\KG}}}-\frac{\sqrt{\alpha^{\KG}}(\lambda^{\KG}+1)}{(\beta^{\KG}+\lambda^{\KG})}\right] 
	\; \sin\iota
	\; \eta^{1/5} \; \frac{\mathcal{M}}{r} \; u \; \cos\Phi\,, 
	\nonumber
	\\
	h_{+}^{\KG} &= 2 \; G_{\AE} \; (1+\cos^2\iota)\; \frac{\mathcal{M}}{r} \; \eta^{1/5} \; u^2 \; \cos(2\Phi)\,,
	\\
	h_{\times}^{\KG} &= 4 \; G_{\AE} \; \cos\iota\; \frac{\mathcal{M}}{r} \; \eta^{1/5} \; u^2 \; \sin(2\Phi)\,,
\end{align}
to leading PN order. Note that the vector modes $h_{\SN}$ and $h_{\SE}$ are now absent~\cite{Blas:2011zd}. Note also that the $h_b$ and $h_l$ modes are linearly dependent, i.e.~their ratio is a function only of the coupling parameters of the theory.

The time-domain response function in khronometric gravity is then
\begin{align}
	h(t)^{\KG} &= A_2^{\KG}\frac{\mathcal{M}}{r}u\left(e^{-2i\Phi+i\Theta}+e^{+2i\Phi-i\Theta} \right)
	\\
	\nonumber
	&+ A_1^{\KG}\bar{\alpha}^{\KG}\frac{\mathcal{M}}{r}\eta^{1/5}u^2(e^{-i\Phi}+e^{i\Phi})\,,
\end{align}
where recall that $\Theta = \tan^{-1}[2F_\times\cos\iota/F_+(1+\cos^2\iota)]$, and interestingly
\begin{align}
	A_2^{\KG} &= A_2^{\EA}\,,
	\\
A_1^{\KG} &= 4 \; G_{\AE} \; (s_1^{\KG}-s_2^{\KG})
 \; \,,
	\\
	\label{eqn:alphakg}
	\bar{\alpha}^{\KG} &= \left[\frac{\sqrt{\alpha^{\KG}}(\lambda^{\KG}+1)}{\beta^{\KG} +\lambda^{\KG}}\left(F_b+2F_l\right)
	 	\right. 
	 	\nonumber \\
		&\left. +\frac{1}{\sqrt{\alpha^{\KG}}}\left(F_b-2F_l\right)\right]\,.
\end{align}
Comparing this to Eqs.~\eqref{eq:A2}--\eqref{eq:alpha} in Einstein-\AE{}ther theory, we see that $A_{1,2}$ are very similar, but $\bar{\alpha}$ is quite different, primarily due to the absence of the vector modes in khronometric gravity. Note also that $A_2^{\KG}$ is precisely what one would find in pure GR in the limit $G_{\AE} \to 1$, while $A_1^{\KG}$ goes to zero as the khronometric coupling parameters go to zero, because then $s_{i}^{\KG}$ vanishes fast enough. In Sec.~\ref{sec:Fisher}, it will become clear why we have defined the amplitude of the $\ell =1$ mode of the response function through the product of $A_{1}^{\KG} \bar{\alpha}^{\KG}$. 

%---------------------------------------------------------------------------------
\subsection{Frequency Domain Response}
We now transform the time-domain response function in khronometric gravity to the frequency domain, again using the SPA.
As in the Einstein-\AE{}ther case, we follow the same reasoning presented in Sec.~\ref{sec:Response}.

The rate of change of the binding energy in khronometric gravity is given by Eq.~\eqref{eqn:edotgen} with
\begin{align}
	\delta_{\dot{E}_{b}}^{\KG} &=
	\frac{7}{4}\eta^{2/5}u^{-2} \dot{E}_{-1\PN}^{\KG} + \dot{E}^{\KG}_{0\PN} \,.
		\nonumber
\end{align}
Taking $(s_1-s_2)$ again as a small parameter, we have that
\begin{align}
	\label{eqn:bm1pnkg}
	\dot{E}_{-1\PN}^{\KG} &\equiv \tilde{\beta}_{-1\PN}^{\KG} = \frac{5}{84}\mathcal{G}(s_1-s_2)^2\sqrt{\alpha^{\KG}}
	\nn \\
	& \times \left[\frac{(\beta^{\KG}-1)(2+\beta^{\KG}+3\lambda^{\KG})}{(\alpha^{\KG}-2)(\beta^{\KG}+\lambda^{\KG})}\right]^{3/2}\,,
	\\
	\label{eqn:b0pnkg}
	\dot{E}_{0\PN}^{\KG} &\equiv \tilde{\beta}_{0\PN}^{\KG} 
	= \mathcal{G} \left(1-\frac{2}{\beta_{\KG}}\right)
	\nn \\
	& \times
	\left(\frac{1}{w_{2}^{\KG}}+\frac{3\alpha_{\KG}(Z_{\KG}-1)^2}{2w_{0}^{\KG}(2-\alpha_{\KG})} + S\mathcal{A}_2 + S^2\mathcal{A}_3\right)
	-1\,,
\end{align}
where $\mathcal{A}_2$ and $\mathcal{A}_3$ are found in Eqs.~(121) and (122) of~\cite{Yagi:2013ava}, and
\begin{align}
	Z_{\KG} = \frac{(\alpha_1^{\ppN}-2\alpha_2^{\ppN})(1-\beta_{\KG})}{3(2\beta_{\KG}-\alpha_{\KG})}\,.
\end{align}
As in the Einstein-\AE{}ther case, the binding energy is not modified to leading, Newtonian order beyond the simple replacement of gravitational constants. To 1PN order, there will khronometric modifications, but these require knowledge of the derivative of the sensitivities, which are not available. Refer to the discussion below Eq.~\eqref{eq:ZEA} for more details.   

Given the above, the rate of change of the orbital frequency in khronometric gravity is given by Eq.~\eqref{eq:Fdot-def} with 
\begin{align}
	\delta_{\dot{F}}^{\KG} &= \frac{7}{4} \eta^{2/5} \tilde{\beta}_{-1\PN}^{\KG} u^{-2} +  \tilde{\beta}^{\KG}_{0\PN}\,.
\end{align}
From these expressions, we can compute the correction to the integrand of the phase $\nu(F)$ to find
\begin{align}
	\delta_{\nu} &= \left(- \frac{5}{48} \eta^{2/5} \tilde{\beta}_{-1\PN}^{\KG} u^{-2} +  \tilde{\beta}^{\KG}_{0\PN} \right) (F\ell-f)\,.
\end{align}

The Fourier response is then again given by Eq.~\eqref{eq:htwidle}, where the Fourier phase in khronometric gravity is
\begin{align}
	\label{eq:phasekg}
	\Psi_{\ell}^{\KG} &= 2\pi ft_c+\Phi_c - \frac{\pi}{4}-\frac{3\ell}{256}u^{-5}_{\ell} \left[1 + {\cal{O}}(c^{-2})\right]
	\nonumber
	\\
	&-\frac{3\ell}{256}u^{-5}_{\ell} \left[\tilde{\beta}_{-1\PN}^{\KG}\eta^{2/5}u_{\ell}^{-2} + \tilde{\beta}_{0\PN}^{\KG} + {\cal{O}}(c^{-2})\right]\,,
\end{align}
and the Fourier amplitudes are
\begin{align}
	\mathcal{A}_1^{\KG} &= \left(\frac{5\pi}{48}\right)^{1/2} A_1^{\KG}\bar{\alpha}^{\KG}\frac{\mathcal{M}^2}{r_{12}}\eta^{1/5}u_1^{-9/2}\,,
	\\
	\mathcal{A}^{\KG}_2 &= \left(\frac{5\pi}{96}\right)^{1/2} A^{\KG}_2\frac{\mathcal{M}^2}{r_{12}}u_2^{-7/2}\,.
\end{align}
As before, the GR prediction for the SPA phase is in the first line of Eq.~\eqref{eq:phasekg}, and it can be taken to as high PN order as one wishes. 

Note that the Fourier response in khronometric gravity is identical to that in Einstein-\AE{}ther theory, up to how the $\bar{\alpha}_{-1,0\PN}$ and $\tilde{\beta}_{-1,0\PN}$ quantities depend on the coupling parameters of the theory.  With this in mind it is obvious that khronometric gravity too can be mapped easily to the ppE framework, via the same mapping given in Eq.~\eqref{eq:ppE}, but of course replacing the $\EA$ labels with the corresponding $\KG$ ones. Similarly, we can decompose the khronometric response function in the same way as in Einstein-\AE{}ther theory through Eq.~\eqref{eq:responsedecomp}.  We find we obtain the precise form of Eqs.~\eqref{eqn:responsedecompea}-\eqref{eq:lastEA}, but with
\begin{align}
	\alpha_v^{\KG} &= 0\,,
	\\
	\alpha_b^{\KG} &= 4G_{\AE}(s_1-s_2)\left[\frac{\sqrt{\alpha^{\KG}}(\lambda^{\KG}+1)}{\beta^{\KG}+\lambda^{\KG}}+\frac{1}{\sqrt{\alpha^{\KG}}}\right]\,,
	\\
	\alpha_l^{\KG} &= 8G_{\AE}(s_1-s_2)\left[\frac{\sqrt{\alpha^{\KG}}(\lambda^{\KG}+1)}{\beta^{\KG}+\lambda^{\KG}}-\frac{1}{\sqrt{\alpha^{\KG}}}\right]\,.
\end{align}

%%%%%%%%%%%%%%%%%%%%%%%%%%%%%%%%
\section{Projected Constraints on Lorentz Violating Gravity with Only Gravitational Waves}
\label{sec:Fisher}

Let us assume that a gravitational wave has been observed by a given detector and that the observation is consistent with GR. Let us then try to estimate the accuracy to which a non-GR deviation can be claimed to be consistent with zero, given statistical uncertainties. We will do so here through a Fisher analysis following the presentation in~\cite{Will:1994fb, cutlerflanagan}. 

\subsection{Basics of a Fisher Analysis}
Let us first define some data analysis quantities that will be necessary for a Fisher analysis. Given a waveform template, $h$, the SNR is defined via
\begin{align}
	\rho(h) &= (h\vert h)^{1/2}\,,
\end{align}
where we have introduced the inner-product between two templates $h_{1}$ and $h_{2}$ via~\cite{cutlerflanagan}
\begin{align}
	(h_1\vert h_2) \equiv 2\int\limits_{0}^{\infty}\frac{\tilde{h}_1^*(f)\tilde{h}_2(f)+\tilde{h}_1(f)\tilde{h}_2^*(f)}{S_n(f)}df\,.
\end{align}
Here, the superscript star stands for complex conjugation, while $S_n(f)$ is a detector's noise spectrum. We will here use the zero-detuned, high-power spectral noise of aLIGO~\cite{mishra}, as well as the projected noise spectrum of LIGOIII~\cite{Adhikari:2013kya}, ET~\cite{mishra, et} and DECIGO~\cite{Cutler:2009qv, Berti:2004bd}. In practice, the lower limit of integration will be set to 10Hz for aLIGO and LIGOIII, 1Hz for ET, and $10^{-3}$Hz for DECIGO, while the upper limit of integration will be set to $2 F_{\rm{ISCO}}$, where $F_{\rm{ISCO}}\equiv 6^{-3/2}/(2 \pi m)$ is the orbital frequency for a particle in a Schwarzschild background at the innermost stable circular orbit. 

We use two modified gravity waveform models in our analysis.  The first one is a \emph{fully-restricted} model, in which we keep up to 3.5PN order terms in the GR phase~\cite{Buonanno:2009zt}, up to 1 PN order terms in the non-GR phase relative to the leading-order correction, and model the Fourier amplitude with only the leading-order PN expression in GR, namely
\begin{align}
	\tilde{h}_{\fr, \MG}(f) &= \mathcal{A} \; f^{-7/6} \; e^{i\Psi_2^{ \MG}}\,,
\end{align}
with $\Psi_2^{ \MG} = \Psi_{\ell=2}^{ \MG}$ and 
\begin{align}
	\Psi_\ell^{ \MG} &= \delta \Psi^{ \MG}_\ell + \frac{3\ell}{256u_\ell^5}\sum\limits_{n=1}^{7}u_\ell^n(p_n^{\PN}+l_n^{\PN}\ln(u_{\ell}))\,.
\end{align}
Here, $p_n^{\PN}$ and $l_n^{\PN}$ are known PN coefficients in GR (see e.g.~\cite{Buonanno:2009zt}), while 
\begin{align}
	\label{eqn:deltapsi}
	\delta\Psi_\ell^{\MG} = -\frac{3\ell}{256}u_\ell^{-5}\left(\tilde{\beta}^{\MG}_{-1\PN}\eta^{2/5}u_\ell^{-2} + \tilde{\beta}^{\MG}_{0\PN} \right)\,,
\end{align}
with $\tilde{\beta}_{\MG} = \tilde{\beta}_{\EA}$ or $\tilde{\beta}_{\KG}$ depending on whether we are studying Einstein-\AE{}ther theory or khronometric gravity. 

The second modified gravity model we will employ is a \emph{semi-restricted} one, in which we keep up to 3.5PN order terms in the GR phase, up to 1 PN order terms in the non-GR phase, and we model the Fourier amplitude as a linear superposition of both GR and non-GR terms to leading PN order, namely
\begin{align}
	\tilde{h}_{\sr, \MG}(f) &= \mathcal{A}f^{-7/6}e^{i\Psi_2^{\MG}}
	\nonumber \\
	 &-\left(\frac{288}{1225}\right)^{1/2} \mathcal{A} \, \bar{\alpha}_{\MG} \, \eta^{1/5}f^{-3/2}e^{i\Psi_1^{\MG}}\,,
\end{align}
where $\Psi_{\ell}^{\MG}$ was given in Eq.~\eqref{eqn:deltapsi}, while $\bar{\alpha}_{\MG} = \bar{\alpha}_{\EA}$ or $\bar{\alpha}_{\KG}$ depending on whether we are studying Einstein-\AE{}ther theory or khronometric gravity [see Eqs.~\eqref{eq:alpha} and~\eqref{eqn:alphakg}, respectively]. Note that the amplitude of the $\ell=1$ mode depends on the coupling parameters of the modified theory only through $\bar{\alpha}_{\MG}$, which is why introduced  correction $\bar{\alpha}_{\EA}$ or $\bar{\alpha}_{\KG}$ in Secs.~\ref{sec:AE-theory} and~\ref{sec:KG-theory}.

Both models assume an \emph{angle-averaged} response function with parameters 
\be
\Theta^{a}_{\fr}=(t_{c}, \phi_{c},{{\cal{A}},\cal{M}},\eta,\tilde{\beta}_{-1\PN,\MG}, \tilde{\beta}_{0\PN, \MG})\,,
\ee
for the fully-restricted model and 
\be
\Theta^{a}_{\sr}=(t_{c}, \phi_{c},{{\cal{A}},\cal{M}},\eta,\bar{\alpha}_{\MG}, \tilde{\beta}_{-1\PN,\MG}, \tilde{\beta}_{0\PN, \MG})\,,
\ee
for the semi-restricted one. In this parameter list, recall that $(t_{c},\phi_{c})$ are the time and phase of coalescence, ${\cal{M}}$ is the chirp mass, and $\eta$ is the symmetric mass ratio. The quantity ${\cal{A}} = (245\pi/6144)^{1/2}\mathcal{M}^{5/6}/r$ is an overall amplitude constant that depends on the chirp mass, distance from the source to the detector and is angle averaged over all sky angles. 

Given a gravitational wave detection, the accuracy to which template parameters can be estimated (in Gaussian and stationary noise) can be approximated via
\begin{align}
	\Delta\Theta^a_{\mbox{\tiny{RMS}}} = \sqrt{\Sigma^{aa}}\,,
\end{align}
where $\Sigma^{ab} = (\Gamma_{ab})^{-1}$ is the inverse of the Fisher matrix,
\begin{align}
	\Gamma_{ab} \equiv \left. \left(\frac{\partial h}{\partial \Theta^a} \right| \frac{\partial h}{\partial\Theta^b}\right)\,,
\end{align}
If the detection is consistent with GR (i.e.~evaluating the Fisher matrix with $\tilde{\beta}_{-1\PN, \MG}, \tilde{\beta}_{0\PN, \MG}$ and $\bar{\alpha}_{\MG}$ all set to zero.), then $\Delta \beta_{\MG}$ and $\Delta \bar{\alpha}_{\MG}$ lead to projected constraints on the coupling parameters of Lorentz-violating gravity. In the next subsections, we will estimate these constraints. 

%--------------------------------------------------------------------------------
\subsection{Projected Constraints on Einstein-\AE{}ther Theory and Khronometric Gravity}
%--------------------------------------------------------------------------------

		\begin{table*}[ht]
			\centering
			\begin{tabular}{c|c|c|c|c|c|c|c|c|c|}
				\cline{2-10}
				 & $m_1$ & $ m_2 $ & $\Delta t_c$ (ms) & $\Delta \Phi_c$ & $\frac{\Delta {\cal{M}}}{{\cal{M}}}$ & $\frac{\Delta \eta}{\eta}$ & $\Delta\bar{\alpha}_{\MG}$ & $\Delta \tilde{\beta}_{-1\PN,\MG}$ & $\Delta \tilde{\beta}_{0\PN,\MG}$\\
				\hline\hline
				 & $1.4 M_\odot$ & $1.4 M_\odot$ & 0.680 &  6.22& $0.225$ & $2.71\times 10^{-2}$ &2.71 &   $5.32\times 10^{-5} $ & 0.389\\
				\cline{2-10}
				aLIGO & $2.0 M_\odot$ & $1.0 M_\odot$ & 0.651 &  7.18 & $0.234$ & $1.85\times 10^{-2}$ &2.77 &  $5.71\times 10^{-5}$ &  0.403\\
				\cline{2-10}
				& $1.8 M_\odot$ & $1.2 M_\odot$ & 0.666 &  6.79 & $0.243$ & $2.08\times 10^{-2}$ &2.73 &  $6.11\times 10^{-5}$ & 0.419\\
				\cline{2-10}
				& $1.4 M_\odot$ & $1.5 M_\odot$ & 0.678 &  6.41 & $0.237$ & $2.30\times 10^{-2}$ &2.71 &  $5.80\times 10^{-5}$ & 0.408\\
				\cline{2-10}\cline{2-10}
				\hline
				\hline
				& $1.4 M_\odot$ & $1.4 M_\odot$ & 0.179  & 1.59 & $5.56\times 10^{-2}$ & $7.07\times 10^{-3}$ & 0.913 & $1.17\times 10^{-5}$& $9.55\times 10^{-2}$  \\
				\cline{2-10}
				LIGOIII & $2.0 M_\odot$ & $1.0 M_\odot$ & 0.168 &  1.78 & $5.67\times 10^{-2}$ & $5.27\times 10^{-3}$ & 0.935 &  $1.24\times 10^{-5}$& $9.76\times 10^{-2}$\\
				\cline{2-10}
					& $1.8 M_\odot$ & $1.2 M_\odot$ & 0.172 &  1.70 & $5.92\times 10^{-2}$ & $6.03\times 10^{-3}$ & 0.921 &  $1.33\times 10^{-5}$& 0.102\\
				\cline{2-10}
					& $1.4 M_\odot$ & $1.5 M_\odot$ & 0.177 &  1.62 & $5.80\times 10^{-2}$ & $6.77 \times 10^{-3}$ & 0.913 &  $1.27\times 10^{-5}$& $9.97\times 10^{-2}$\\
				\hline
				\hline
				& $1.4 M_\odot$ & $1.4 M_\odot$ & $3.96\times 10^{-2}$ & $0.123$ & $1.91\times 10^{-3}$ & $1.20\times 10^{-3}$ & 0.196 &   $8.60\times 10^{-8} $ & $3.23\times 10^{-3}$\\
				\cline{2-10}
				ET & $2.0 M_\odot$ & $1.0 M_\odot$ & $3.94\times 10^{-2}$ &  $0.125$ & $1.88\times 10^{-3}$ & $1.04\times 10^{-3}$ & 0.201 &  $9.01\times 10^{-8}$ & $3.19\times 10^{-3}$\\
				\cline{2-10}
					& $1.8 M_\odot$ & $1.2 M_\odot$ & $3.95\times 10^{-2}$ &  $0.123$ & $2.00\times 10^{-3}$ & $1.14\times 10^{-3}$ & 0.198 &  $9.71\times 10^{-8}$ & $3.38\times 10^{-3}$\\
				\cline{2-10}
					& $1.4 M_\odot$ & $1.5 M_\odot$ & $3.96\times 10^{-2}$ &  $0.123$ & $1.98\times 10^{-3}$ & $1.19 \times 10^{-3}$ & 0.196 &  $9.32\times 10^{-8}$& $3.35\times 10^{-3}$\\
				\cline{2-10}
				\hline
				\hline
				& $1.4 M_\odot$ & $1.4 M_\odot$ & 1.80 & 0.295 & $9.50\times 10^{-5}$ & $9.25\times 10^{-4}$ & $5.42\times 10^{-2}$ &   $2.92\times 10^{-10} $& $1.59\times 10^{-4}$\\
				\cline{2-10}
				DECIGO & $2.0 M_\odot$ & $1.0 M_\odot$ & 1.80 &  0.295 & $8.15\times 10^{-5}$ & $8.68\times 10^{-4}$ & $5.55\times 10^{-2}$ &  $3.05\times 10^{-10}$ & $1.37\times 10^{-4}$\\
				\cline{2-10}
				& $1.8 M_\odot$ & $1.2 M_\odot$ & 1.78 &  0.289 & $9.24\times 10^{-5}$ & $8.98\times 10^{-4}$ & $5.45\times 10^{-2}$ &  $3.07\times 10^{-10}$ & $1.55\times 10^{-4}$\\
				\cline{2-10}
				& $1.4 M_\odot$ & $1.5 M_\odot$ & 1.79 &  0.290 & $9.66\times 10^{-5}$ & $9.19 \times 10^{-4}$ & $5.41\times 10^{-2}$ &  $3.00\times 10^{-10}$ & $1.62\times 10^{-4}$\\
				\cline{2-10}
			\end{tabular}
			\caption{\label{tbl:Estimates} Parameter estimation accuracy using aLIGO (top), LIGOIII (top center), ET (lower center), and DECIGO (bottom).  The simulated signals for all three detectors lie at a luminosity distance of approximately $270$Mpc, which corresponds SNR's of $\rho_{\mbox{\tiny{aLIGO}}} = 10$, $\rho_{\mbox{\tiny{LIGOIII}}}=30$, $\rho_{\mbox{\tiny{ET}}} = 130$, $\rho_{\mbox{\tiny{DECIGO}}} = 140$.   Notice that in general, ET is able to estimate parameters to an order of magnitude greater accuracy than aLIGO for a source at the same distance. In particular, note that ET can constrain both $\tilde{\beta}_{-1\PN, \MG}$ and $\tilde{\beta}_{0\PN, \MG}$ more than two orders of magnitude more stringently than aLIGO, while LIGOIII constraints generally lie between those obtained with the other two detectors. Note also that the bounds calculated on $\Delta\bar{\alpha}$ can be obtained only in the semi-restricted model. Finally, note that these are the constraints produced in \textit{both} Einstein-\AE{}ther theory and in khronometric theory, since the waveforms are functionally the same with differences only in how $\tilde{\beta}_{\MG}$ is related to the coupling parameters of the theory. }
		\end{table*}

Using the method described above, we can calculate projected constraints on the modified theory template parameters $(\tilde{\beta}_{-1\PN,\MG},\tilde{\beta}_{0\PN,\MG})$, as well as the accuracy to which other system parameters can be measured. This is shown in Table~\ref{tbl:Estimates}, where we have assumed gravitational wave observations consistent with GR from the quasi-circular inspiral of non-spinning NSs with second- and third-generation detectors. We place the binaries at a fixed luminosity distance ($D_{L} = 270 \; {\rm{Mpc}}$) and average over all sky positions, leading to SNRs of $10$, $30$, $130$ and $140$ with aLIGO, LIGOIII, ET and DECIGO respectively. The results reported in Table~\ref{tbl:Estimates} scale with the reciprocal of the SNR. 

Let us concentrate on the columns of Table~\ref{tbl:Estimates} that present the projected bounds on the modified gravity theory template parameters. Note that the projected bounds on $\bar{\alpha}_{\MG}$ can only be calculated using the semi-restricted model, since the fully restricted one does not have $\bar{\alpha}_{\MG}$ as a template parameter. Note also that the columns that give bounds on $\tilde{\beta}_{-1\PN, \MG}$ and $\tilde{\beta}_{0\PN, \MG}$ give exactly the same constraints irrespective of using the full or the restricted model, since the additional term in the semi-restricted model scales with $\bar{\alpha}_{\MG}$. Obviously, the constraints on the modified gravity template parameters improve with SNR, but observe that they are almost insensitive to the NS masses.

Are the projected gravitational wave constraints on $(\tilde{\beta}_{-1\PN,\MG},\tilde{\beta}_{0\PN,\MG},\tilde{\beta}_{0\PN,\MG},\bar{\alpha}_{\MG})$ presented in Table~\ref{tbl:Estimates} competitive with current binary pulsar constraints on the coupling parameters? Saturating the latter and using the weak-field limit approximation for the sensitivities leads to $(\tilde{\beta}_{-1\PN,\MG}^{\rm{bin.pul}},\tilde{\beta}_{0\PN,\MG}^{\rm{bin.pul}},\bar{\alpha}_{\MG}^{\rm{bin.pul}}) \lesssim (10^{-4},10^{-3},10^{-2})$. Note that $\bar{\alpha}_{\MG}^{\rm{bin.pul}}$ is smaller than the projected constraints quoted in Table~\ref{tbl:Estimates}, while $(\tilde{\beta}_{-1\PN,\MG}^{\rm{bin.pul}},\tilde{\beta}_{0\PN,\MG}^{\rm{bin.pul}})$ are comparable or even less strict than those in Table~\ref{tbl:Estimates}. This suggests that gravitational waves will not be able to produce competitive bounds on the coupling parameters through modifications to the amplitude ($\bar{\alpha}_{\MG}$), but may be able to compete via measurements of the phase modifications $\tilde{\beta}_{0\PN,\MG}$ and $\tilde{\beta}_{-1\PN,\MG}$. Henceforth, we concentrate on the latter only.

Given the results in Table~\ref{tbl:Estimates}, we can now map the constraints on $(\tilde{\beta}_{-1\PN, \MG},\tilde{\beta}_{0\PN, \MG})$ to constraints on the coupling parameters of Einstein-\AE{}ther and khronometric gravity via
\begin{align}
\label{eq:inequality1}
\tilde{\beta}_{-1\PN,\EA}(c_{+},c_{-}) &\leq \tilde{\beta}_{-1\PN,\MG}\,,
\\
\label{eq:inequality1a}
\tilde{\beta}_{0\PN,\EA}(c_{+},c_{-}) &\leq \tilde{\beta}_{0\PN,\MG}\,,
\end{align}
and
\begin{align}
\tilde{\beta}_{-1\PN,\KG}(\lambda_{\KG},\beta_{\KG}) &\leq \tilde{\beta}_{-1\PN,\MG}\,,
\\
\tilde{\beta}_{0\PN,\KG}(\lambda_{\KG},\beta_{\KG}) &\leq \tilde{\beta}_{0\PN,\MG}\,.
\label{eq:inequality2}
\end{align}
The right-hand sides are given in Table~\ref{tbl:Estimates}, which have a weak dependence on the NS masses. To be conservative, we here pick component masses $(m_{1},m_{2}) = (1.2,1.8) M_{\odot}$, but we have checked that picking other masses, i.e.~other rows in Table~\ref{tbl:Estimates}, leads to negligible effects on the coupling parameter constraints. 
The left-hand sides are given in Eqs.~\eqref{eqn:bm1pn}--\eqref{eqn:b0pn} and~\eqref{eqn:bm1pnkg}--\eqref{eqn:b0pnkg}, which depend on all coupling parameters and on the NS sensitivities. We set $\alpha_{1,2}^{\ppN} = 0$ in both theories, since $(c_+, c_-,\lambda_{\KG},\beta_{\KG}) \gg \alpha_{1,2}^{\ppN}$, given current constraints on the latter. 
Once the sensitivities are written in terms of the coupling parameters, each inequality produces a distinct contour in the $(c_{+},c_{-})$ and $(\lambda_{\KG},\beta_{\KG})$ space, and the intersection of these contours serves as a projected bound on the coupling parameters.  

% % % % % % % % % % % % % % % % %
\begin{figure*}[ht]
\includegraphics[width=8.75cm,clip=true]{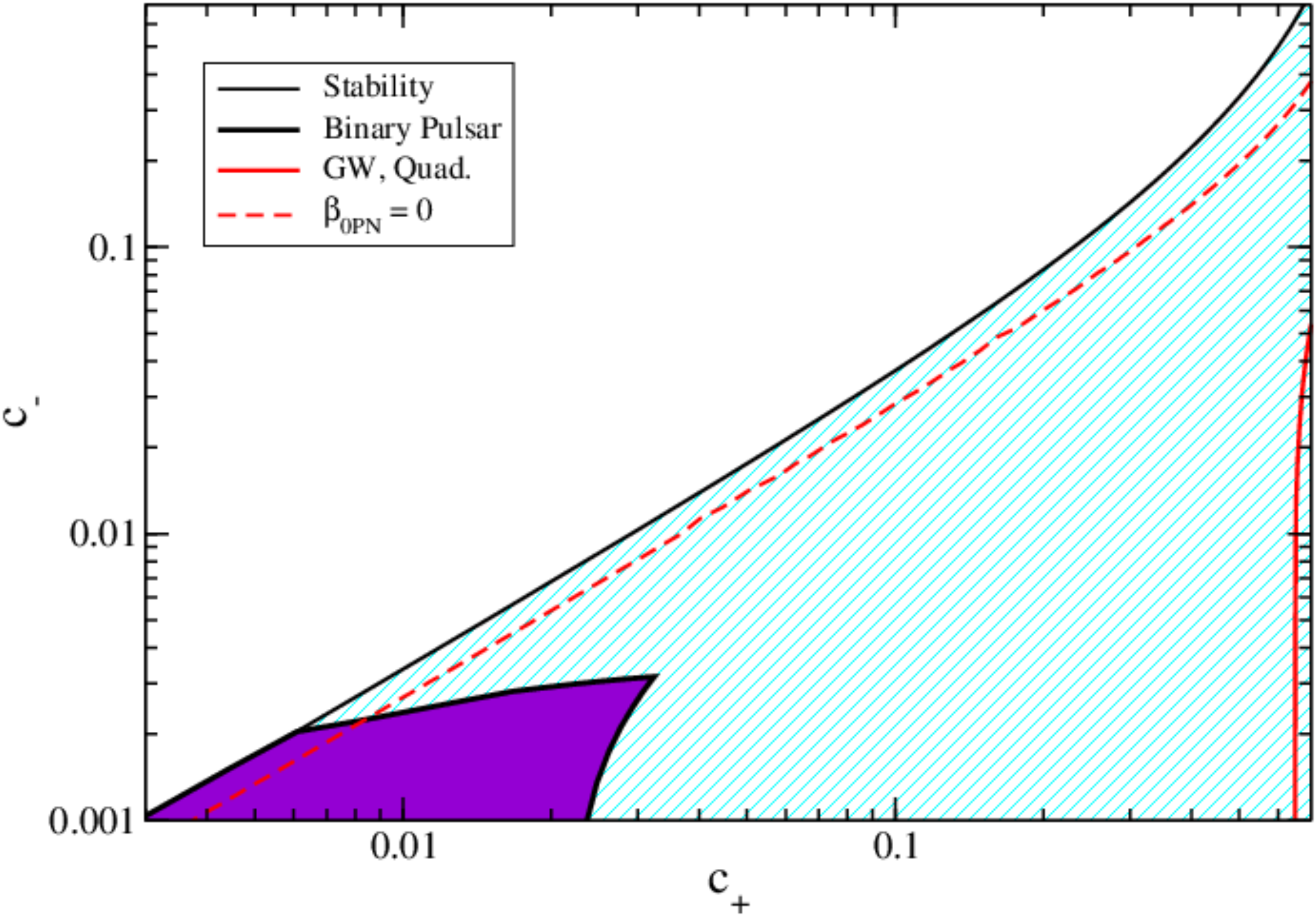}
\includegraphics[width=8.75cm,clip=true]{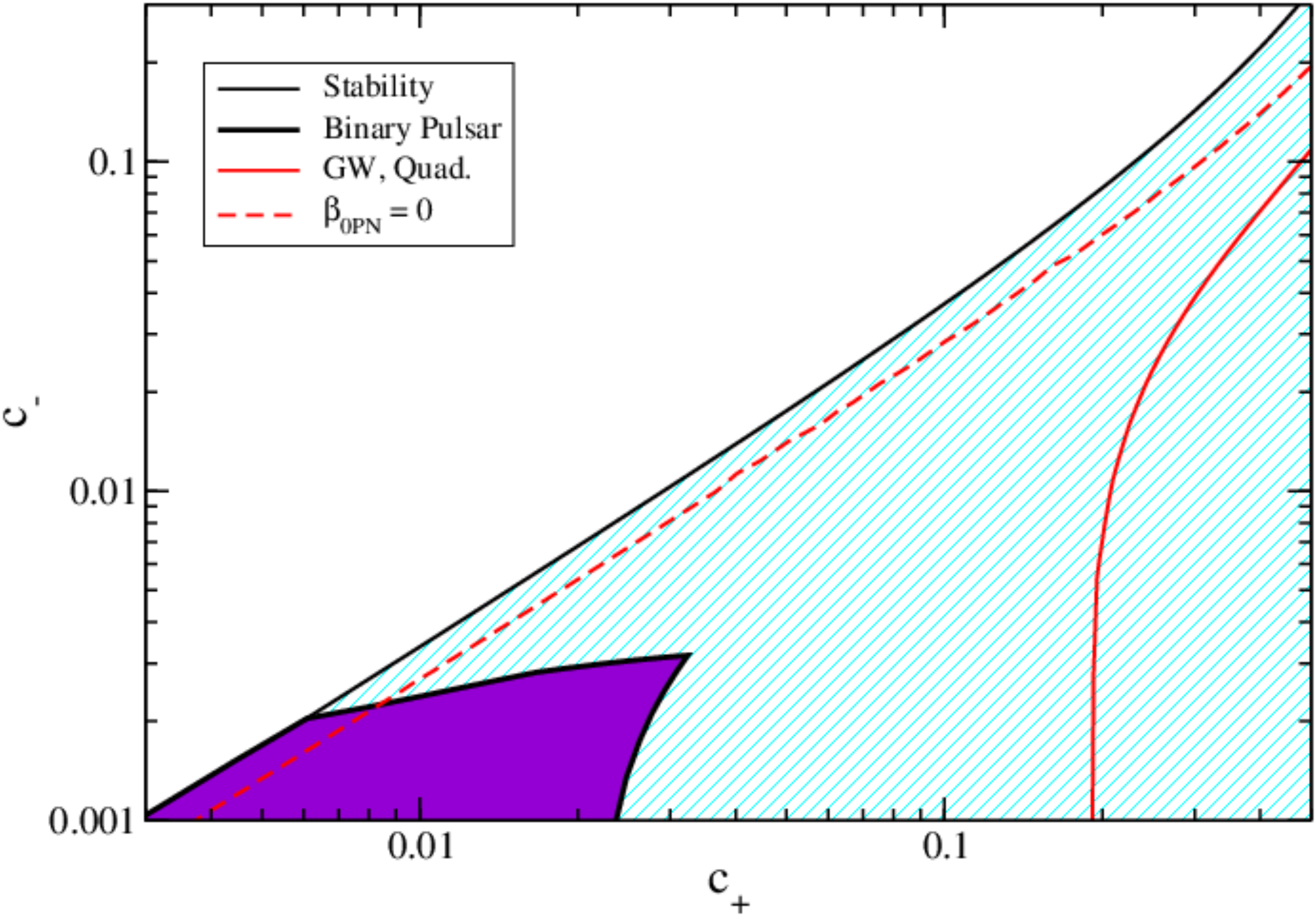} 
\includegraphics[width=8.75cm,clip=true]{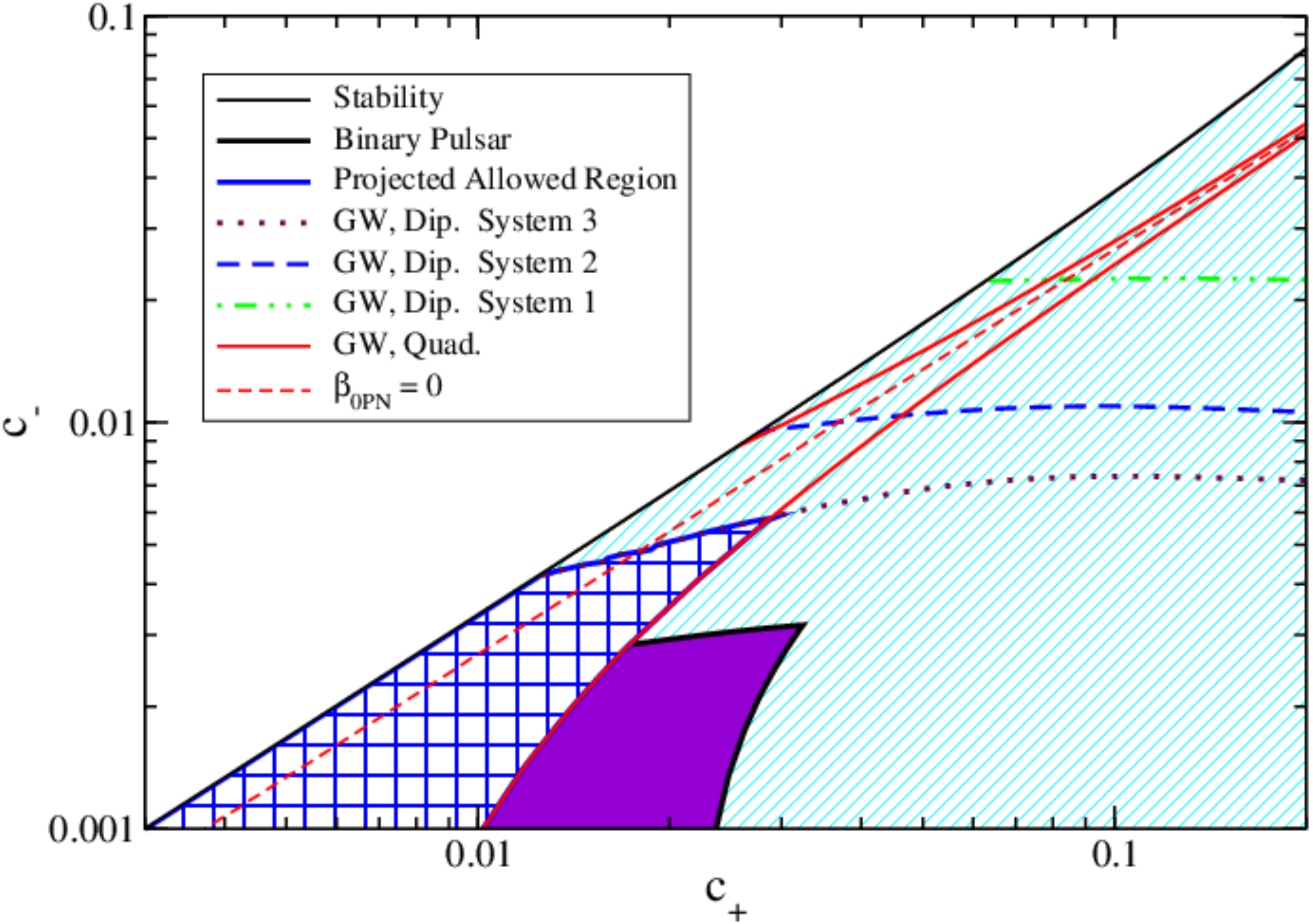}
\includegraphics[width=8.75cm,clip=true]{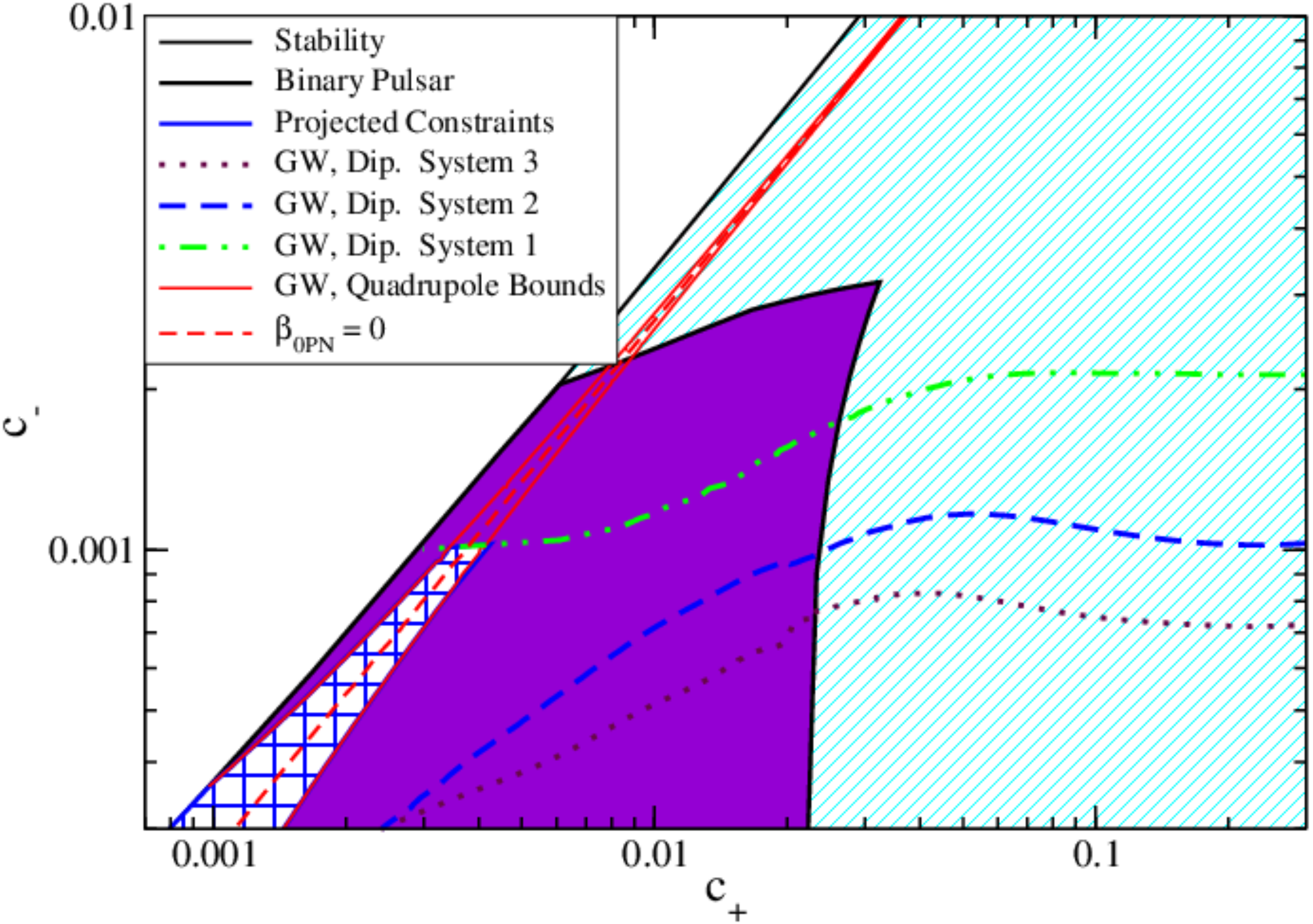}
\caption{\label{fig:ConstraintsET} Projected constraints on the $(c_+,c_-)$ space, given a gravitational wave detection by aLIGO, LIGOIII, DECIGO and ET (clockwise from top left) consistent with GR. We inject a simulated signal produced by the quasi-circular inspiral of non-spinning NSs at a luminosity distance of $270 \; {\rm{Mpc}}$. The contours are produced by finding the values of $(c_+,c_-)$ that satisfy Eqs.~\eqref{eq:inequality1} and~\eqref{eq:inequality1a}. Thus, the region below the contours are the allowed regions for $c_+$ and $c_-$.  One cannot place constraints from $\bar{\beta}_{-1\PN, \EA}$ with aLIGO or LIGOIII because the measurement errors on the masses are too large due to degeneracy with $\beaz$. The dark shaded (purple) region is the allowed region from binary pulsar observations, as calculated in \cite{Yagi:2013qpa,Yagi:2013ava}.  Note that the gravitational wave projected constraints are less strict by orders of magnitude than those calculated with binary pulsar observations. }
\end{figure*}
% % % % % % % % % % % % % % % % % % %

An important caveat is here in order: the mapping of constraints on $\tilde{\beta}_{-1\PN,\MG}$ to constraints on the coupling parameters of the theory is impossible for systems of exactly equal masses. This is because $\tilde{\beta}_{-1\PN,\EA}$ and $\tilde{\beta}_{-1\PN,\KG}$ depend on $(s_{1} - s_{2})^{2}$ [see Eqs.~\eqref{eqn:bm1pn} and~\eqref{eqn:bm1pnkg}]. For any given value of the coupling parameters, if the NS masses are the same, then the sensitivities are also the same (assuming the same equation of state for each NS component) and the effect of dipole radiation disappears.  Henceforth, we will consider only unequal mass NS binaries. 

Before we can map projected constraints on $(\tilde{\beta}_{-1\PN,\MG},\tilde{\beta}_{0\PN,\MG})$ to projected constraints on the coupling parameters, we must express the sensitivities as functions of the latter. This has been done in~\cite{Yagi:2013ava} through a fitting function of the form  
\begin{align}
	s_{A}^{\EA,\KG} = \sum\limits_{\ell, m, n=0}^{2} c_{\ell,m,n}^{\EA,\KG} \kappa_{1}^{\ell} \kappa_{2}^{m} C_{*,A}^n\,,
\end{align} 
where $C_{*,A}$ is the stellar compactness of the Ath star (the ratio of the individual stars mass to its radius), $(\kappa_{1},\kappa_{2}) = (c_{+},c_{-})$ and $(\kappa_{1},\kappa_{2}) = (\beta_{\KG},\lambda_{\KG})$ in Einstein-\AE{}ther and khronometric gravity respectively, and where the $c_{\ell, m, n}^{\EA,\KG}$ constants are given in Table I of~\cite{Yagi:2013ava}. The fit found in~\cite{Yagi:2013ava}, however, is not very useful in this paper. This is because the former is valid only in a very specific region of coupling parameters, which was relevant in~\cite{Yagi:2013ava} but is not in this paper. Moreover, the fit does not lead to zero sensitivities in the limit as $(c_+,c_-) \to 0$, because Ref.~\cite{Yagi:2013ava} chose $\alpha_{1,2}^{\ppN}$ by saturating Solar System constraints. 

We thus create new fitting functions for $s_{A}$ by re-running the code developed in~\cite{Yagi:2013ava}. We will require that the sensitivities go to zero in the limit as the coupling parameters go to zero, by setting $\alpha_{1,2}^{\ppN} = 0$. We will further restrict attention to the region in $(c_+, c_-)$ and $(\beta_{\KG},\lambda_{\KG})$ that is relevant to this paper, as shown e.g.~in Fig.~\ref{fig:Constraints}. Finally, as we will be interested only in a discrete set of compactnesses, we create separate fits for each compactness used. This allows us to use lower-order polynomials in the fitting function and have the same fitting accuracy. 

As described above, the sensitivities depend on the compactness of the star, which, given a fixed mass, depends only on its radius, and thus, on its equation of state. As the NS equation of state is still unknown, we will consider three representative systems with difference radii, and thus compactnesses: 
\begin{enumerate}
	\item {\emph{System 1:}} $C_{*,1} = 0.12$ and $C_{*,2} = 0.13$.
	\item {\emph{System 2:}} $C_{*,1} = 0.08$ and $C_{*,2} = 0.15$.
	\item {\emph{System 3:}} $C_{*,1} = 0.07$ and $C_{*,2} = 0.14$.
\end{enumerate}  
The different compactnesses act as a proxy for the variability of the equation of state, and together with the fitting functions, completely determine the sensitivities as functions of the coupling parameters only. 

Using these representative systems, we are able to project bounds on $(c_{+},c_{-})$ and $(\lambda_{\KG},\beta_{\KG})$ using Eqs.~\eqref{eq:inequality1} and~\eqref{eq:inequality2}. Figure~\ref{fig:ConstraintsET} presents these bounds for Einstein-\AE{}ther theory. In all four panels, the dark (purple) shaded region marks the allowed region of coupling parameter space, as calculated in~\cite{Yagi:2013qpa,Yagi:2013ava} from binary pulsar observations. As these are currently the most strict constraints on the coupling parameters of Einstein-\AE{}ther theory, any bounds which can be placed which exclude parts of this region increase our ability to constrain the theory.  All four panels also contain a stability region, shown through a light (cyan) shading and delimited by a thin (black) stability line. This region corresponds to values of $c_+$ and $c_-$ for which the speeds of propagation of the modes equal or exceed the speed of light, thus satisfying the stability criterion discussed in earlier sections. 

The top-left (top-right) panel depicts the bounds placed using aLIGO (LIGOIII).  The solid red curve marks the outer edge of the region allowed using constraints on $\tilde{\beta}_{0\PN, \EA}$; all points to the left of the curve cannot be ruled out by our analysis.  Recall that $\tilde{\beta}_{0\PN, \EA}$ depends on the sensitivities very weakly, only through $S {\cal{A}}_{2} + S^{2} {\cal{A}}_{3}$ [see e.g.~Eq.~\eqref{eqn:b0pn}], which allows us to use a single curve for all three representative systems.  The dotted red curve marks the line in coupling parameter space where $\tilde{\beta}_{0\PN, \EA} = 0$. Thus, if the coupling parameters of Einstein-\AE{}ther theory lied on this line, one would never be able to constrain the theory with observations of the 0PN correction to the gravitational wave phase. The projected bounds derived from constraints on $\tilde{\beta}_{0\PN, \EA}$ form a ``track" surrounding the $\tilde{\beta}_{0\PN, \EA} = 0$ line.  However, for aLIGO and LIGOIII, the leftmost bound on the track lies outside of the region allowed by stability and is thus not shown explicitly in the figure.  Observe that the solid curve in the LIGOIII case is farther to the left than in the aLIGO case, but neither of these bounds is competitive with current binary pulsar constraints.

Observe also that neither of these top panels presents bounds on $(c_{+},c_{-})$ from constraints on $\tilde{\beta}_{-1\PN, \EA}$.  This is because $\tilde{\beta}_{-1\PN, \EA}$ depends strongly on $(s_1-s_2)^2$, which in turn is proportional to the difference in compactness of the two stars.  Thus, for two equal mass (and equal compactness) stars, $\tilde{\beta}_{-1\PN, \EA} = 0$. We consider here only NS binaries with unequal masses, and thus, compactnesses for which $\tilde{\beta}_{-1\PN,\EA}$ is small but non-zero. However, for both aLIGO and LIGOIII, the error in recovery of the chirp mass is quite large because this parameter enters at 0PN order and is thus degenerate with $\tilde{\beta}_{0\PN, \EA}$.  For instance, aLIGO has a 20\% error in chirp mass recovery for the systems examined in Table~\ref{tbl:Estimates}.  For any of the three cases examined here, we would be unable to ensure from a detection that the masses of the two stars were not actually equal. The most conservative bounds then come from the case where $C_{*, 1}=C_{*, 2}$, in which case $s_1-s_2 = 0$ and therefore \textit{all} values of the coupling parameters satisfy the bounds on $\tilde{\beta}_{-1\PN, \EA}$ and constraints come only from bounds on $\tilde{\beta}_{0\PN,\EA}$. 

The bottom left (right) panel presents the constraints on Einstein-\AE{}ther theory projected for ET (DECIGO).  The region between the two solid red curves marks the space allowed after placing constraints on $\tilde{\beta}_{0\PN, \EA}$. Observe that both for ET and DECIGO, the gravitational wave constraints can improve upon constraints from binary pulsar analysis alone. Unlike in the aLIGO or LIGOIII case, we can now also place constraints on Einstein-\AE{}ther theory from bounds on $\beta_{-1\PN,\EA}$. This is because ET and DECIGO would be able to extract the component masses to sufficient accuracy that the equal mass case could be excluded to high confidence for the systems investigated, thus ruling out the possibility that $\beta_{-1\PN,\EA}$ could vanish due to $s_{1} = s_{2}$. Bounds extracted from $\beta_{-1\PN,\EA}$ are mostly-horizontal lines (dotted, dashed, dash-dot-dotted in the figure) for the three representative systems. As expected, the strongest constraints come from systems with the most dissimilar compactnesses (the most unequal mass systems).
The overlap of the constraint regions produced by bounds on $\beta_{0\PN, \EA}$ and $\beta_{-1\PN, \EA}$ lead to the \emph{combined constraints} on Einstein-\AE{}ther theory. In other words, given a gravitational wave detection with ET or DECIGO, one would be able to constrain the coupling parameters of Einstein-\AE{}ther theory to be in the overlap of the $\beta_{0\PN, \EA}$ region and the $\beta_{-1\PN, \EA}$ of the system detected. As an example, this is shown as a checkered (blue) region. Observe that these combined constraints can notably improve upon binary pulsar bounds. 

%--------------------------------------------------------------------------------
% % % % % % % % % % % % % % % % %
\begin{figure*}[ht]
\includegraphics[width=8.75cm,clip=true]{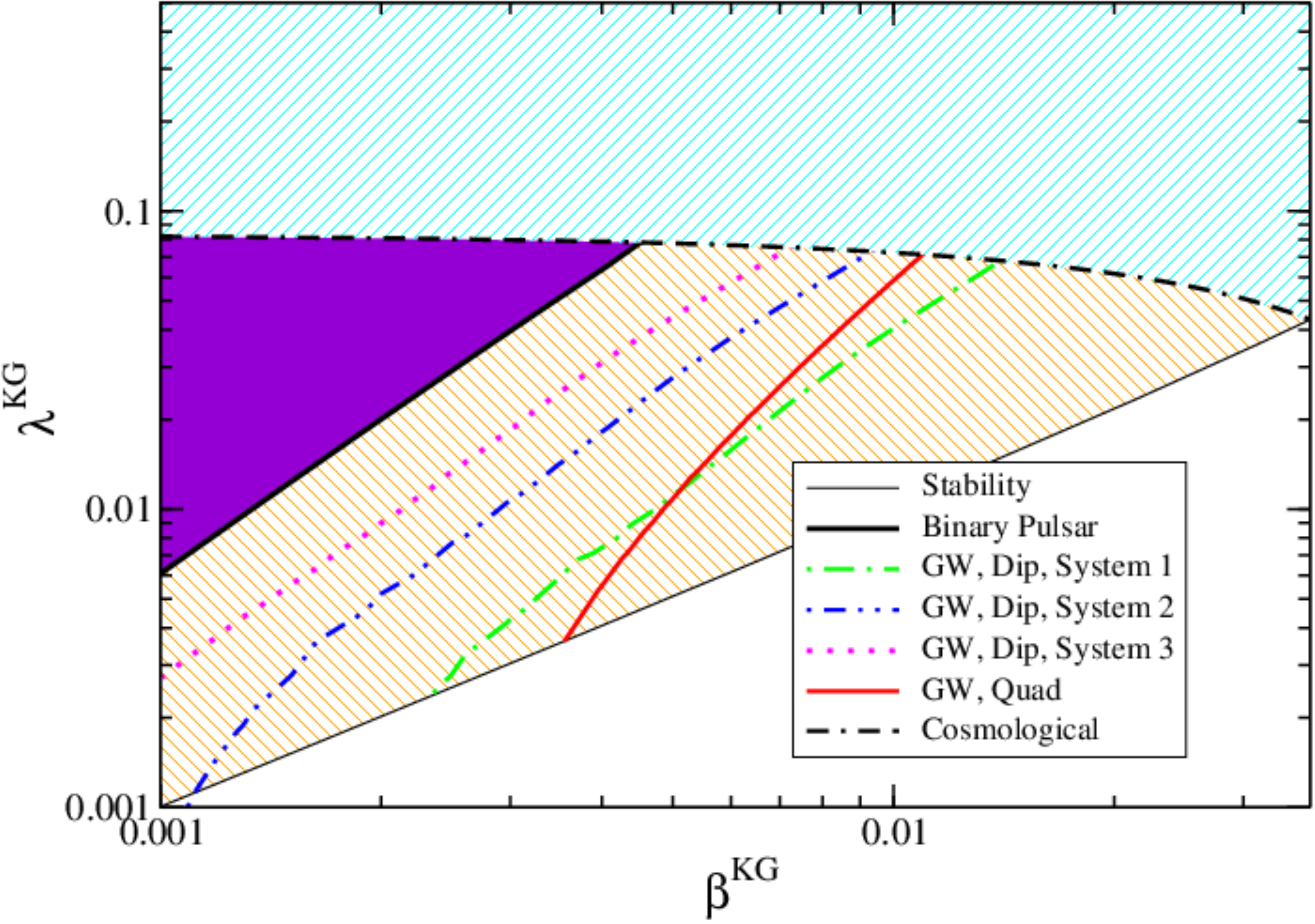} 
\includegraphics[width=8.75cm,clip=true]{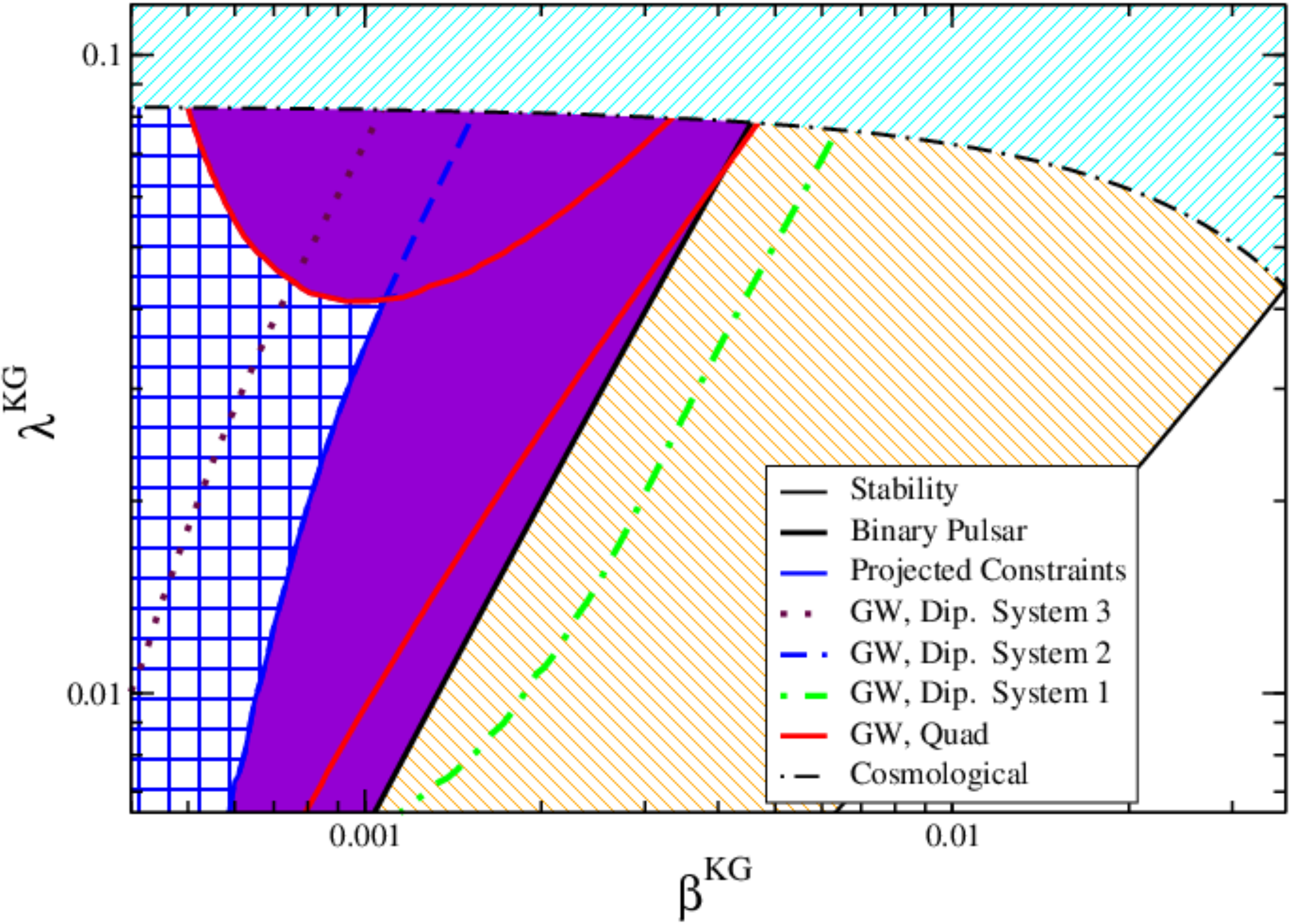}
\caption{\label{fig:ConstraintsKG}   Projected constraints on $\beta^{\KG}$ and $\lambda{^{\KG}}$ via a gravitational wave detection by 3rd generation detectors ET (Left) and DECIGO (Right), consistent with GR.  The contours are produced by finding the values of $\beta_{\KG}$ and $\lambda{\KG}$ such that the bounds on $\bkgz$ and $\bkgo$ are met exactly.  Thus, the region above (or bounded by) the contours are the allowed regions for $\beta_{\KG}$ and $\lambda_{\KG}$.  The dark (purple) shaded region is the allowed region from binary pulsar observations, as calculated in \cite{Yagi:2013qpa,Yagi:2013ava}, the light (cyan) shaded region is that allowed by stability, and the mid (orange) shaded region is the allowed region from cosmological constraints.  Note that the constraints producible via second generation gravitational wave  detection are entirely outside the region allowed by requiring stability and cosmological constraints.}
\end{figure*}
% % % % % % % % % % % % % % % % % % %
Not surprisingly, our findings are similar in khronometric theory (Fig.~\ref{fig:ConstraintsKG}). Due to our choice of notation and the way khronometric gravity effects correct GR, the results of our Fisher analysis are \textit{identical} to those produced on Einstein-\AE{}ther theory, except for the way $\tilde{\beta}_{-1\PN, \MG}$ and $\tilde{\beta}_{0\PN, \MG}$ depend on $\lambda_{\KG}$ and $\beta_{\KG}$. Due to the combination of stability and cosmological constraints, we find that the constraints produced on khronometric theory are only competitive when using ET and DECIGO; the aLIGO and LIGOIII bounds allow the entire region which is not ruled out by stability and cosmological observations.  

Figure~\ref{fig:ConstraintsKG} shows the projected bounds on the coupling parameters in khronometric gravity given a gravitational wave detection consistent with GR.  In both panels, the dark (purple) shaded region depicts the allowed values of the coupling parameters after constraints from binary pulsar observations~\cite{Yagi:2013qpa,Yagi:2013ava} are imposed.  The light (cyan) section marks the region where stability constraints are satisfied, and the medium (orange) region is the section where both stability and cosmological constraints are satisfied. As in the Einstein-\AE{}ther case, the bounds on the coupling constants obtained with an aLIGO or a  LIGOIII detection are less strict than stability constraints, and are we will not report them here.

The left (right) panel depicts the projected constraints assuming a detection with ET (DECIGO).  We present both the bounds given by $\bkgz$ (solid red curve) as well as $\bkgo$ for the three representative systems described earlier.  Since $\bkgz$ again depends weakly on compactness, we only present a single curve for each detector.  Note that the contours produced by ET approach but do not surpass current binary pulsar bounds, while a detection with DECIGO could narrow the allowed region of coupling parameter space for khronometric gravity. The overlap of the constraints obtained with $\bkgz$ and $\bkgo$ form the combined projected bound, shown here as a checkered, shaded region using system II.

% % % % % % % % % % % % % % % % % % % % % % % % % %
\section{Projected Constraints on Lorentz Violating Gravity with Coincident Gravitational Waves and Electromagnetic Observations}
\label{sec:gwem}

Although using only gravitational wave observations from NS binaries with second-generation detectors will be insufficient to place bounds on the coupling parameters of Lorentz violating gravity theories that are stronger than current constraints, we can gain much additional information if a coincident electromagnetic counterpart is observed.  Such an event could occur if, for example, binary NS mergers are progenitors to  short gamma ray bursts, and one were to occur with such a geometry that Earth lied in the line of sight. Similarly, one could place constraints given a supernovae that resulted in a neutrino burst, provided it occurred close enough to Earth for detection. While such coincident events are far from likely, a single coincidence would be enough to place very interesting constraints. 

%------------------------------------------------------------------------
\subsection{Constraints from Times of Arrival}

Given a gravitational wave observation with an associated electromagnetic counterpart, one could use the time of arrivals to infer the propagation speed of photons and gravitons, given the distance to the source~\cite{Nishizawa:2014zna}. In the case of a short gamma-ray burst, the latter can be estimated from the gravitational wave observation itself during the binary NS inspiral. In the case of a supernova, the distance could be estimated electromagnetically, provided the explosion occurs close enough for gravitational wave detection (i.e.~roughly within our galaxy). Of course, the statistical error in the determination of the distance can be non-negligible, but even folding this error into account, we will see that constraints on the coupling parameters from coincident detections can be very powerful. 

A slight complication is that there is an \emph{intrinsic} time delay between photon or neutrino emission and gravitational wave emission \cite{Baret:2011tk}. The quantum-mechanical processes that produce neutrino and photon emission may precede or follow the core bounce or the NS merger, while the gravitational wave emission is centered tightly around this event~\cite{Nishizawa:2014zna}.  However, these delays have been very well constrained \cite{Baret:2011tk, Ott:2012kr, Marek:2008qi}.  Section~II of~\cite{Nishizawa:2014zna} derives an explicit relation between the intrinsic time delay of emission and the bounds on the difference of the propagating speeds of gravitational waves and photons.  Below we will utilize the results of this analysis to constrain the effects of Lorentz violating gravity.

Keeping this modeling caveat in mind, let us define the speed of gravitons via
\begin{align}
	w_2 = c(1-\delta_g)\,,
\end{align}
where $\delta_{G}$ is a function of the coupling constants of the theory only and, in pure GR, $\delta_g = 0$. Given a binary NS merger at a luminosity distance of approximately 200Mpc with a GR signal and an electromagnetic counterpart, a conservative bound on $\delta_g$ was estimated to be~\cite{Nishizawa:2014zna}
\begin{align}
	\label{eq:deltag}
	|\delta_g| \lessapprox 10^{-14}\,.
\end{align} 
The approximately less or equal sign is included because there are slight differences in the bounds when one uses different models for the photon and neutrino time delays.  We use here the most conservative estimates presented in \cite{Nishizawa:2014zna}.

Given such a constraint on $\delta_{g}$, we can then constrain Lorentz violating gravity, because generically the propagation speed of the tensor modes is modified in such theories. In Einstein-\AE{}ther theory, the propagation speed is given by $w_{2}^{\EA}/c = (1-c_+)^{-1/2}$ [see Eq.~\eqref{eqn:aespeed}], while in khronometric gravity it is given by $w_{2}^{\KG}/c = (1-\beta_{\KG})^{-1/2}$ [see Eq.~\eqref{eq:prop-speed-KG}], where we have reinserted the factors of $c$ for clarity. Then, from Eq.~\eqref{eq:deltag} we automatically obtain
\begin{align}
	c_+ \lessapprox 10^{-14}\,, \qquad \beta_{\KG} \lessapprox 10^{-14}\,.
\end{align}
These projected bounds are much more stringent than any other constraint by over 10 orders of magnitude. 

The bounds on $c_{+}$ and $\beta_{\KG}$ presented above, of course, depend on our accuracy to constrain $\delta_{g}$, which in turn is limited by systematic errors induced by the intrinsic time of emission of electromagnetic radiation~\cite{Nishizawa:2014zna}. Let us assume the massive, electromagnetic particle travel time $T_{\rm{em}}$ and the graviton travel times $T_{g}$ are roughly $T_{\rm{em},g} \sim (D/c) \gamma_{{\rm{em}},g}$, where $D$ is the distance to the source, while $\gamma_{\rm{em},g} = 1 - v_{\rm{em},g}/c$. The time lag between their arrival is then simply $\Delta T = (D/c) (\gamma_{\rm{em}} - \gamma_{g})$. Let us assume now that we have measured no time lag, but our measurement has a systematic uncertainty due to our intrinsic ignorance for the time of emission of electromagnetic radiation relative to the emission of gravitational radiation: $(\Delta T)_{\rm m} = 0 + \Delta \tau_{\rm{int}}$. Then, we can place the constraint
\be
\label{eq:coinc}
\left|\gamma_{\rm{em}} - \gamma_{g}\right| < \frac{c}{D} \; \Delta \tau_{\rm int}
\ee
on $\gamma_{g}$. How well we can constrain $\gamma_{g}$ then depends on $\gamma_{\rm{em}}$ and on $\Delta \tau_{\rm int}$. 

\begin{figure}[ht]
\includegraphics[width=8.75cm,clip=true]{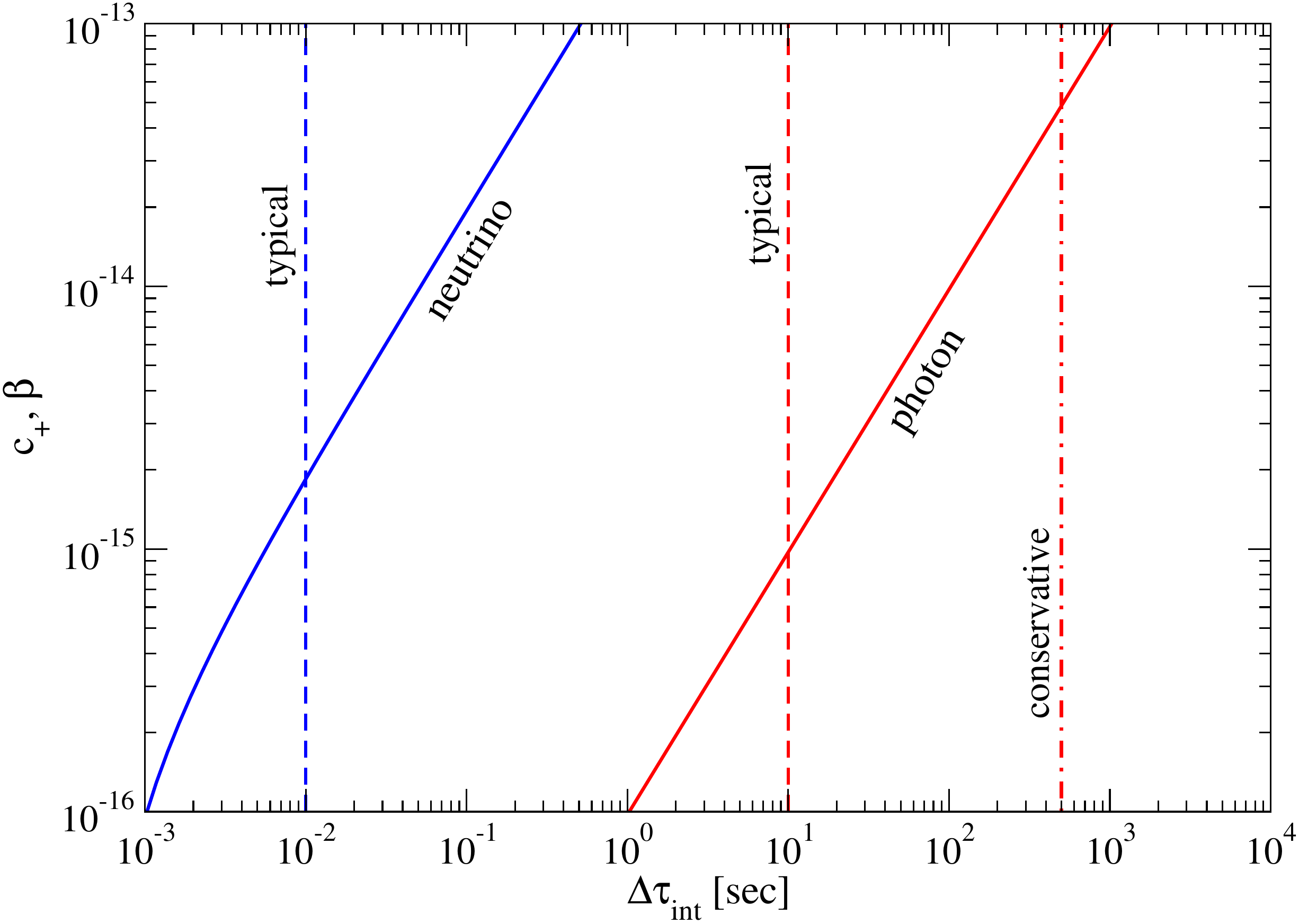}
\caption{\label{fig:gwem} Projected constraints on $c_+$ and $\beta^{\KG}$, assuming a coincident gravitational wave and electromagnetic observation. The vertical (dashed) lines show conservative and typical estimates for the intrinsic uncertainty in the time of arrival of electromagnetic radiation. The solid lines are the solutions to Eq.~\eqref{eq:coinc} for $c_{+}$ and $\beta^{\KG}$. The intersection of the solid and dashed lines show the projected constraints.}
\end{figure}

Figure~\ref{fig:gwem} presents these constraints assuming a photon or neutrino detection that is coincident with a gravitational wave detection. For Einstein-\AE{}ther theory and khronometric gravity, we have $\gamma_{g}^{\EA} \sim -c_{+}/2 $ and $\gamma_{G}^{\KG} = -\beta_{\KG}/2$. In the neutrino case, we set $\gamma_{\nu} = m_{\nu}^{2} c^{4}/(2 E_{\nu}^{2})$, with $m_{\nu} = 0.1$ eV and $E_{\nu} = 10$ MeV, while in the photon case, $\gamma_{\gamma} = 0$. We assume $D = 100$ kpc and $200$ Mpc for the neutrino (from a supernova) and the photon (from a short gamma-ray burst) cases respectively, with measurement errors much smaller than those incurred by $\Delta \tau_{\rm{int}}$. For the latter, we present conservative and ``typical'' estimates from~\cite{Nishizawa:2014zna} (vertical dashed lines in Fig.~\ref{fig:gwem}). The solid lines in Fig.~\ref{fig:gwem} are then the solution to Eq.~\eqref{eq:coinc} for $c_{+}$ and $\beta_{\KG}$. Thus, the intersection of the solid lines with the vertical dashed lines gives the constraints that could be placed on $c_+$ and $\beta^{\KG}$.

Although given a coincident electromagnetic and gravitational wave observation we can place stringent constraints on $c_{+}$ and $\beta_{\KG}$, the lack of a coincident observation provides no information. The latter can be explained by a vast set of reasons that have nothing to do with GR modifications, such as large misalignment angles for short gamma-ray bursts, the presence of dust between the source and Earth, uncertainties in the sky location, etc. Of course, if the speed of propagation of gravitational waves was sufficiently different from $c$, one would also lose the electromagnetic-gravitational association. However, one would associate the lack of coincidence with astrophysical effects, rather than a modification to GR. 

Given this, there is a vast range in coupling parameter space that one could \emph{not} constrain at all with the method described above.   In practice, a time delay between electromagnetic and gravitational signals of more than a few days would probably be enough to lose confidence in any type of coincidence. One can then ask how large would the coupling parameters of the theory have to be for no coincident events (events with time delays smaller than a few days) to ever occur. For a source at 100Mpc, one finds that this occurs when $(c_+, \beta_{\KG}) \gtrsim 10^{-10}$ in Einstein-\AE{}ther theory or khronometric gravity. From binary pulsar and gravitational wave bounds, we can constrain these coupling parameters to be $\lessapprox 10^{-2}$.  Thus, one would not be able to constrain Lorentz violating gravity for coupling parameters in the range $(10^{-10} \lesssim  c_+\lessapprox 10^{-2})$ and $(10^{-10} \lesssim \beta_{\KG} \lessapprox 10^{-2})$.

\subsection{Recalculating Gravitational Wave Constraints Given a Coincident Detection}

Given a coincident detection, one can effectively set $c_{+}=0$ in Einstein-\AE{}ther theory and $\beta_{\KG} = 0$ in khronometric gravity, and then study what constraints can be placed on the other coupling parameters given additional, non-coincident gravitational wave observations. We will further set $ \alpha_{2}^{\ppN} = 0$, as this parameter has already been constrained strongly to be $\alpha_{2}^{\ppN} \lessapprox 10^{-7}$ from Solar System experiments and $\alpha_{2}^{\ppN} \lessapprox 10^{-9}$ from isolated pulsar observations~\cite{Shao:2013wga}.

Let us first focus on Einstein-\AE{}ther theory. When $c_+ = 0$, one finds $c_1 = c_-/2$ and $c_3 = -c_-/2$, and Eq.~\eqref{eq:alpha2-EA} reduces to
\be
\alpha_{2}^{\ppN,\EA} = \frac{c_{14} (2 c_{14} c_2 + c_{14} - c_2)}{c_2 (2-c_{14})}\,.
\ee
Notice that one can set $\alpha_{2}^{\ppN,\EA} = 0$ in two ways, namely, (i) $c_2 = c_{14}/(1-2 c_{14})$ and (ii) $c_{14}=0$. In the first case, one can solve Eq.~\eqref{eq:alpha1-EA} for $c_4$ and finds $c_4 = -(\alpha_1^{\ppN,\EA} + 2 c_-)/4$, which in turn leads to $c_2 = -\alpha_1^{\ppN,\EA}/(2 \alpha_1^{\ppN,\EA}+4)$. Therefore, one is left with $(c_-, \alpha_1^{\ppN,\EA})$ as two independent coupling parameters. Alternatively, in the second case, one finds $\alpha_1^{\ppN,\EA} = 0$ and $c_4 = - c_-/2$, while $c_2$ is undetermined. Therefore, one is left with $(c_-, c_2)$ as two independent parameters. 

Let us first focus on the first parameter set, namely taking $(c_-, \alpha_1^{\ppN,\EA})$ as our two independent parameters. Before we can place any types of constraints, we must find the sensitivities as a function of $(\alpha_{1}^{\ppN,\EA},c_{-})$. We calculated the latter numerically following the analysis in~\cite{Yagi:2013ava} and found that as long as $c_{+} \ll c_{-}$, then the sensitivities depend \emph{only} on $\alpha_{1}^{\ppN,\EA}$. In fact, we find that the sensitivities are extremely well approximated by the weak-field expression of~\cite{Foster:2007gr}, namely Eq.~\eqref{eq:wf-s}. Taking the same general expressions for the modifications to the dipolar and quadrupolar radiation derived earlier in Sec.~\ref{sec:AE-theory}, but now taking the limit $c_+=0=\alpha_2^{\ppN,\EA}$, we re-calculate the new allowed regions in $(c_-,\alpha_1^{\ppN, \EA})$ space, given the bounds produced in Sec.~\ref{sec:Fisher}.

\begin{figure}[ht]
	\includegraphics[width=8.75cm,clip=true]{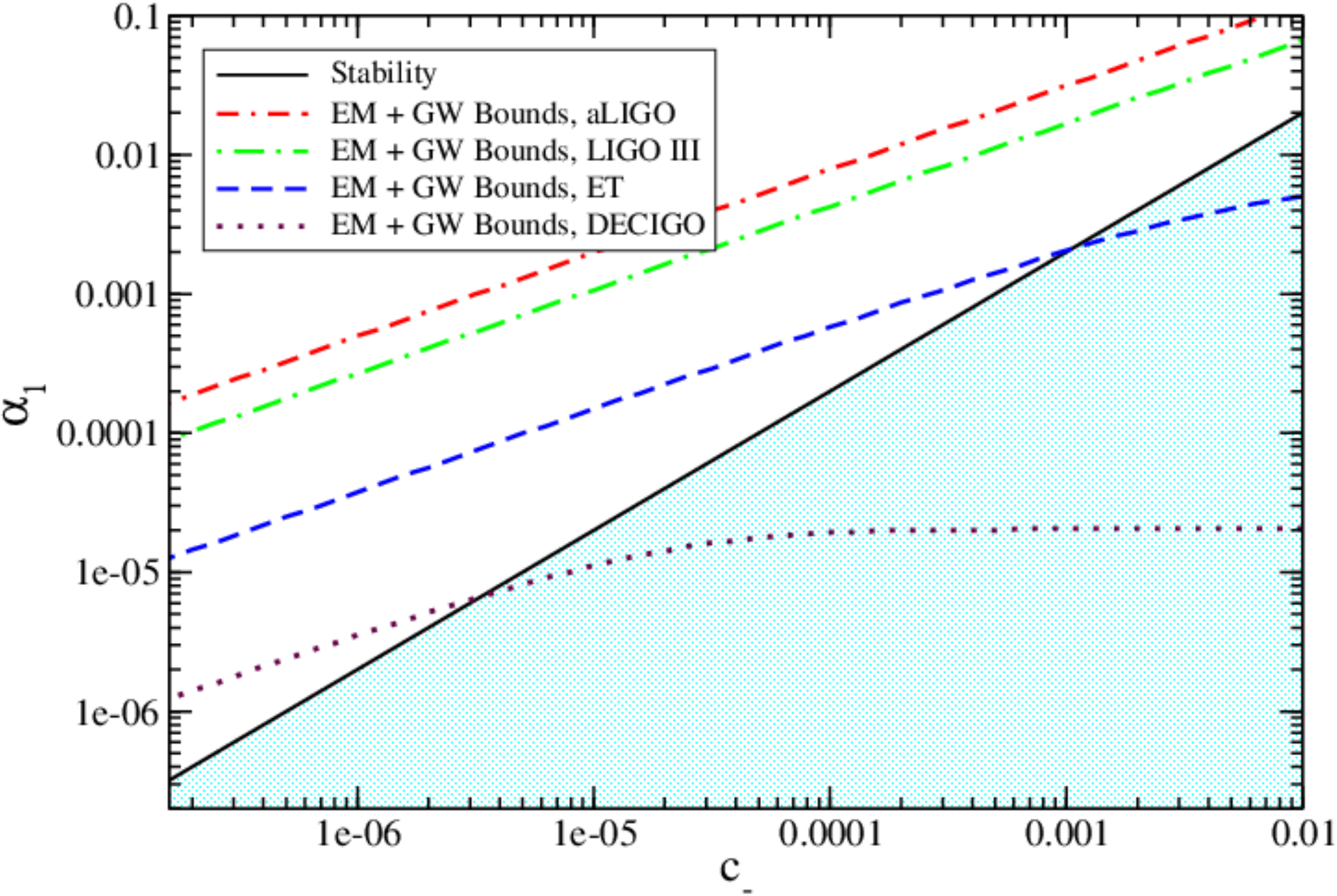}
	\caption{\label{fig:gwembounds}  Bounds on $(\alpha_{1}^{\ppN,\EA},c_{-})$, given a constraint on the speed of gravity through a coincident gravitational wave-electromagnetic observation.  Note that in the region for $c_-$ allowed by stability, the constraints on $\alpha_1^{\ppN,\EA}$ are approximately two orders of magnitude weaker than current Solar System constraints ($\alpha_1^{\ppN,\EA} < 10^{-4}$), and of comparable size to the bounds on $\alpha_1^{\ppN,\EA}$ found in~\cite{Shao:2012eg}.  For third generation detectors, such as ET, the constraints become one order of magnitude stricter than current Solar System constraints.}
\end{figure}
Figure~\ref{fig:gwembounds} shows the bounds on $c_-$ and $\alpha_1^{\ppN,\EA}$ we would be able to place now. Notice that with these new bounds, given a bound on $c_-$, we can constrain $\alpha_1^{\ppN,\EA} \lessapprox \mathcal{O}(c_-)$.  On the other hand, in contrast to the previous cases, we now are not able to place any strong bounds on $c_-$, as the constraint curves are nearly horizontal. As can be seen in Fig.~\ref{fig:Constraints}, the most strict bounds on $c_-$ as $c_+$ becomes small come from stability constraints.  However, as $\alpha_1^{\ppN,\EA}$ and $\alpha_2^{\ppN,\EA}$ cease to be negligible in comparison to $c_\pm$, the stability conditions are modified from what is shown in Fig.~\ref{fig:Constraints}, which is actually only valid when $c_{\pm} \gg \alpha_{1,2}$.  These new stability curves can be found simply by solving the inequalities $w_1 >1$ and $w_0 >1$ for $\alpha_2^{\ppN,\EA}$ in terms of $c_-$.

Let us now consider the second choice of the parameter set, namely taking $(c_-,c_2)$ as our two independent parameters. When one sets $c_+=0$ and expands $\tilde{\beta}_{-1\PN}^{\EA}$ and $\tilde{\beta}_{0\PN}^{\EA}$ in Eqs.~\eqref{eqn:bm1pn} and~\eqref{eqn:b0pn} around $c_{14}=0$, one finds that $\tilde{\beta}_{-1\PN}^{\EA} = \mathcal{O}(c_{14}^2)$ and $\tilde{\beta}_{0\PN}^{\EA} = \mathcal{O}(c_{14})$. This means that the Einstein-\AE{}ther corrections to the gravitational waveform at -1PN and 0PN order vanish in the limit $c_{14} \to 0$ and one cannot place any constraints on the theory from gravitational wave observations.

While in principle we could perform a similar calculation in khronometric theory, this would be fruitless because the most conservative bounds on $\lambda^{\KG}$ come from the case in which $\alpha^{\KG}=2\beta^{\KG}$.  Then, when we impose the constraint $\beta^{\KG}\leq 10^{-14}$, we automatically also have $\alpha^{\KG} \lesssim 10^{-14}$, and we are merely focussing on the far left portion of the bounds already constructed. From Fig.~\ref{fig:ConstraintsKG}, this means that at most $\lambda_{\KG} \in (0, 0.1)$.  Indeed, when we perform this calculation, we find that these are exactly the bounds produced.

% % % % % % % % % % % % % % % % % % % % % % % % % % %
\subsection{Recalculating Binary Pulsar Constraints Given a Coincident Detection}
\label{sec:BinaryPulsar}

Given a coincident detection, we can also recalculate binary pulsar constraints~\cite{Yagi:2013ava} under the condition $c_{+}=0$ in Einstein-\AE{}ther theory and $\beta_{\KG} = 0$ in khronometric gravity.  The rate of change of the orbital period of such a binary in Einstein-\AE{}ther theory is given by~\cite{Yagi:2013ava}
\begin{align}
\label{eq:PdotP}
	\frac{\dot{P}}{P}=\left({\frac{\dot{P}}{P}}\right)_{\GR}\mathcal{A}\,,
\end{align}
where,
\begin{align}
	\label{eqn:acurly}
	\mathcal{A} \equiv 1 + \tilde{\beta}_{0\PN,\MG}+\frac{5}{32} \left(1-\frac{c_{14}}{2}\right) \left(\frac{P_b}{2\pi m}\right)^{2/3}\tilde{\beta}_{-1\PN,\MG}\,,
\end{align}
where, clearly, in pure GR $\mathcal{A}=1$. We then translate observed values for $\dot{P}/\dot{P}_{\GR}$, where $\dot{P}_{\GR}$ is the rate of change of the orbital period predicted by GR, directly into an allowed region for $\mathcal{A}$.

\begin{figure}[t]
	\includegraphics[width=8cm,clip=true]{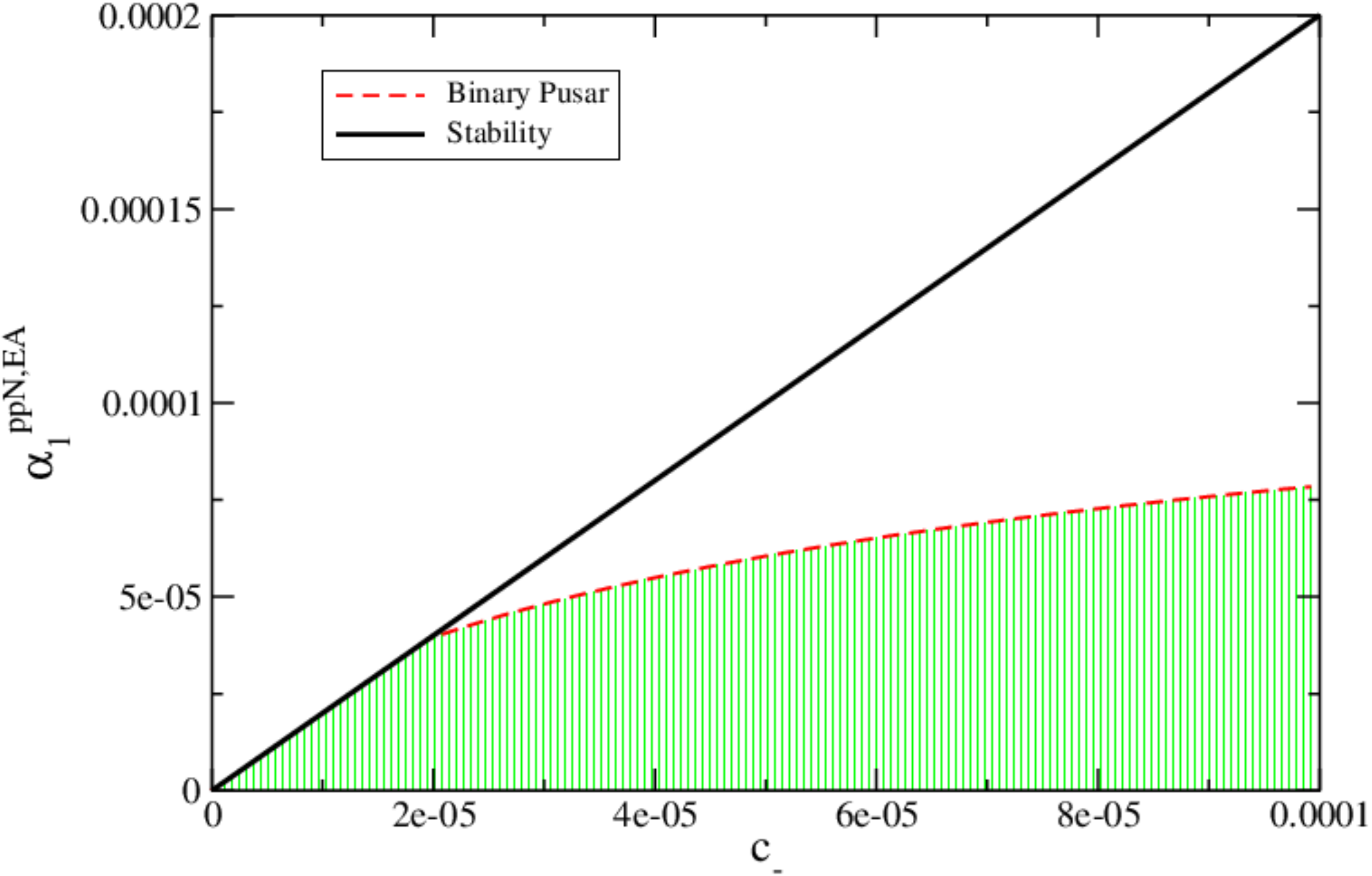}
	\caption{\label{fig:bpbounds}  In the figure above, we show the allowed regions in parameter space for the coupling parameters $c_-$ and $\alpha_1^{\ppN}$ in Einstein-\AE{}ther theory based on binary pulsar observations, in the case of a restriction on the speed of gravity through coincident gravitational wave/electromagnetic detection.  For our analysis, we used observational bounds on the values of $\dot{P}/\dot{P}_{\GR}$ to bound the magnitude of $\mathcal{A}$.  We chose a representative case, PSR J1141-6545~\cite{Bhat:2008ck}.  We also considered pulsars: PSRJ03848+0432~\cite{Antoniadis:2013pzd}, PSR J0737-3039~\cite{Kramer:2006nb}, and PSR J1738-0333~\cite{Freire:2012mg}, and found comparable results.}
\end{figure}

Let us first consider the case where one uses $(c_-,\alpha_{1}^{\ppN,\EA})$ as two independent parameters. Figure~\ref{fig:bpbounds} shows the projected allowed regions in $(c_-, \alpha_{1}^{\ppN,\EA})$ given the four separate binary pulsar observations PSR J1141-6545~\cite{Bhat:2008ck}, PSRJ03848+0432~\cite{Antoniadis:2013pzd}, PSR J0737-3039~\cite{Kramer:2006nb}, and PSR J1738-0333~\cite{Freire:2012mg} found to be consistent with GR. We see that the $(\alpha_{1}^{\ppN,\EA},c_{-})$ region allowed is not heavily restricted. In particular, observe that $\alpha_1^{\ppN, \EA} \lessapprox 10^{-4}$, a bound comparable with Solar System constraints, but no additional bounds could be placed on $c_-$. Of course, these projected bounds on $(\alpha_{1}^{\ppN,\EA},c_{-})$, are possible only through direct measurement of gravitational radiation coincident with an electromagnetic signal. 

Let us now consider the case where one uses $(c_{-},c_{2})$ as two independent parameters. As in the previous subsection, the Einstein-\AE{}ther corrections to the orbital decay rate of a binary pulsar in Eq.~\eqref{eq:PdotP} vanish when $c_+ = 0 = c_{14}$. Therefore, once more, one cannot place any constraints on the theory when one takes $(c_-,c_2)$ as the two independent parameters.

Once again, we do not calculate constraints for khronometric theory in this case, as in the limit as $\beta^{\KG}=0$ is already shown in~\cite{Yagi:2013ava}.  In summary, like before, the most conservative constraints require $\alpha^{\KG} = 2\beta^{\KG}$, which constrains $\lambda^{\KG} \leq 0.1$.  

%%%%%%%%%%%%%%%%%%%%%%%%%%%%%%%%%%%%%%%%%%%%%%%%%%%%
\section{Conclusions}
\label{sec:Conclusions}

The observation of gravitational waves has great potential as a way to test and constrain modified gravity theories.  One such modification is the violation of Lorentz-symmetry in the gravitational sector. In this paper, we studied whether gravitational waves could place meaningful constraints on Lorentz-violating gravity by considering two particular theories: Einstein-\AE{}ther theory and khronometric gravity. 

We first calculated waveform templates for the time-domain and SPA frequency-domain response function, assuming an impinging gravitational wave produced in the late, quasi-inspiral of NS binaries.  We found that the modification to the waveform template can be easily mapped to the ppE framework through a -1PN and a 0PN term in the waveform phase, relative to the leading-order GR quadrupole term. We then performed a Fisher analysis to determine how well Lorentz-violating gravity could be constrained given a gravitational detection consistent with GR. We found that second-generation detectors, such as aLIGO, will not be able to place constraints that are more stringent than current binary pulsar ones, given a NS inspiral detection. Third-generation detectors, such as LIGOIII, ET and DECIGO, could potentially place more stringent constraints. 

We then examined the possibility of placing constraints on Lorentz-violating gravity given a coincident gravitational wave and electromagnetic observation, for example from a short gamma ray burst or a supernova explosion. A single coincident event would allow us to place an incredibly stringent constraint on the speed of gravity, relative to the speed of light. This translates into constraints on the coupling parameters of Einstein-\AE{}ther theory and khronometric gravity that are at least $10$ orders of magnitude more strict than current binary pulsar constraints. 

We found that the sensitivities are almost $c_-$ independent when $c_+$ is small, and possible future work could be aimed at understanding such behavior better. Since Foster already showed that the sensitivities do not depend on $c_-$ in the weak-field limit~\cite{Foster:2007gr}, one possible avenue is to extend such an analysis to higher PN order and derive PN corrections to the sensitivities.

Other possibilities for future work include reworking the accuracy to which Lorentz-violating gravity can be constrained with a Bayesian analysis that uses Markov-Chain Monte-Carlo techniques. The Fisher results presented here can be interpreted as best-case, projected constraints, as they are valid only for high SNR events, assuming Gaussian and stationary noise. A more realistic model-selection study would probably lead to more pessimistic constraints. One could do such an analysis for coincident events, as isolated gravitational wave observations with second-generation detectors are not likely to lead to interesting constraints on Lorentz-violating gravity. 

Another possibility for future work would be examining the gravitational wave signals produced by black hole binaries or black hole-neutron star binaries.  For NS binaries, the $-1$PN dipole correction to the GW phase is suppressed by the square of the difference of the sensitivities, while the $0$PN term is partially degenerate with the chirp mass. For mixed binaries, however, the $-1$PN dipole term should be dominant and may lead to much more interesting constraints. However, for such an analysis one would have to first compute the sensitivities in Lorentz-violating gravity for black holes, a calculation that has not yet been carried out. 

%%%%%%%%%%%%%%%%%%%%%%%%%%%%%%%%%%%%%%%%%%%%%%%%%%%%
\section{Acknowledgements}
The authors would like to thank Diego Blas, Neil Cornish, Ted Jacobson and Takahiro Tanaka for discussions.  NY acknowledges support from NSF grant PHY-1114374 and CAREER Award PHY-1250636 as well as support provided by the National Aeronautics and Space Administration from grant NNX11AI49G, under sub-award 00001944. DH would like to acknowledge support from Montana State University Undergraduate Scholars program, as well as the Montana Space Grant Consortium. Some calculations used the computer algebra-systems \textsc{MAPLE}, in combination with the \textsc{GRTENSORII} package~\cite{grtensor}.

\bibliography{master}
\end{document}